\documentclass[11pt]{article}
\usepackage[margin=1truein]{geometry}

\usepackage{setspace}
\usepackage{lscape}
\usepackage{pdflscape}

\usepackage{amsmath,amssymb,amsthm} 

\usepackage{wrapfig} 

\usepackage{mathrsfs} 
\usepackage{url} 
\usepackage{latexsym} 
\usepackage{bm} 

\usepackage{mathtools}

\usepackage{booktabs,multirow}
\usepackage{array}
\newcolumntype{P}[1]{>{\centering\arraybackslash}p{#1}}
\usepackage{diagbox}

\usepackage{algorithm}
\usepackage[noend]{algpseudocode}

\algnewcommand{\Break}{\textbf{break}}
\algnewcommand{\algorithmicgoto}{\textbf{go to} Line}
\algnewcommand{\Goto}[1]{\algorithmicgoto~\ref{#1}}
\algblockdefx{MRepeat}{EndRepeat}{\textbf{repeat}}{}
\algnotext{EndRepeat}

\usepackage[sort&compress,numbers]{natbib}

\usepackage{thmtools,thm-restate}

\usepackage{hyperref}
\usepackage{cleveref}
\expandafter\def\csname ver@etex.sty\endcsname{3000/12/31} 
\usepackage{autonum} 
\makeatletter
\autonum@generatePatchedReferenceCSL{Cref} 
\makeatother

\usepackage{tikz,tikz-3dplot}
\usetikzlibrary{positioning,shapes,shapes.callouts,shapes.multipart,arrows.meta,fit,calc}
\usepackage{pgfplots}
\pgfplotsset{compat=newest}
\usepackage{pxpgfmark} 
\usetikzlibrary{bayesnet} 

\usepackage{multicol}
\usepackage{wasysym} 
\usepackage{adjustbox}
\usepackage{xparse} 
\usepackage{pbox} 
\usepackage{authblk} 
\usepackage{xspace} 

\usepackage[subrefformat=parens]{subcaption}
\usepackage[inline]{enumitem} 
\hypersetup{
  colorlinks,
  linkcolor={green!35!black},
  citecolor={green!35!black},
  urlcolor={blue!50!black}
}

\DeclareMathOperator*{\argmax}{argmax}
\DeclareMathOperator*{\argmin}{argmin}

\DeclareMathOperator{\EE}{\mathbb{E}}

\DeclarePairedDelimiter\ceil{\lceil}{\rceil}

\DeclarePairedDelimiterX\inner[2]{\langle}{\rangle}{{#1},{#2}}

\DeclarePairedDelimiter\set{\{}{\}}
\DeclarePairedDelimiter\prn{(}{)}
\DeclarePairedDelimiter\bra{[}{]}
\DeclarePairedDelimiterX\Set[2]{\{}{\}}{\mspace{2mu}{#1}\;\delimsize|\;{#2}\mspace{2mu}}
\DeclarePairedDelimiterX\Prn[2]{(}{)}{\mspace{2mu}{#1}\;\delimsize|\;{#2}\mspace{2mu}}
\DeclarePairedDelimiterX\Bra[2]{[}{]}{\mspace{2mu}{#1}\;\delimsize|\;{#2}\mspace{2mu}}

\newcommand{\Z}{\mathbb Z}
\newcommand{\R}{\mathbb R}

\newcommand{\st}{\mathrm{s.t.}}

\renewcommand{\epsilon}{\varepsilon}

\NewDocumentCommand{\exsub}{s m O{} m}{%
  \IfBooleanT{#1}{\EE_{#2}\nolimits\bra*{#4}}%
  \IfBooleanF{#1}{\EE_{#2}\nolimits\bra[#3]{#4}}%
}

\usepackage{CJKutf8}

\newcommand{\mathInd}{\hphantom{{}={}}}

\declaretheoremstyle[
shaded={bgcolor=gray!15},
]{thmsty}

\declaretheorem[
  name=Theorem,
  refname={Theorem,Theorems},
  style=thmsty,
]{theorem}

\declaretheorem[
  name=Lemma,
  refname={Lemma,Lemmas},
  style=thmsty,
]{lemma}
\declaretheorem[
  name=Definition,
  refname={Definition,Definitions},
  style=thmsty,
]{definition}

\crefname{algorithm}{Algorithm}{Algorithms}
\crefname{line}{Line}{Lines}
\crefname{section}{Section}{Sections}
\crefname{appendix}{Appendix}{Appendices}
\crefname{table}{Table}{Tables}
\crefname{figure}{Figure}{Figures}
\crefname{equation}{}{}
\Crefname{equation}{Eq.}{Eqs.}

\setlist[itemize]{
  topsep=0.4\baselineskip,
  itemsep=0\baselineskip,
  leftmargin=1.5em,
}

\setlist[enumerate]{
  font=\upshape,
  label=(\alph*),
  ref=(\alph*),
  topsep=0.4\baselineskip,
  itemsep=0\baselineskip,
  leftmargin=2em,
}

\newlist{enuminasm}{enumerate}{1} 
\setlist[enuminasm]{
  font=\upshape,
  label=(\alph*),
  ref=\theassumption(\alph*),
  topsep=0.4\baselineskip,
  itemsep=0\baselineskip,
  leftmargin=2em,
}
\crefalias{enuminasmi}{assumption}

\newlist{enuminthm}{enumerate}{1}
\setlist[enuminthm]{
  font=\upshape,
  label=(\alph*),
  ref=\thetheorem(\alph*),
  topsep=0.4\baselineskip,
  itemsep=0\baselineskip,
  leftmargin=2em,
}
\crefalias{enuminthmi}{theorem}

\newlist{enuminlem}{enumerate}{1}
\setlist[enuminlem]{
  font=\upshape,
  label=(\alph*),
  ref=\thelemma(\alph*),
  topsep=0.4\baselineskip,
  itemsep=0\baselineskip,
  leftmargin=2em,
}
\crefalias{enuminlemi}{lemma}

\newcommand{\email}[1]{\href{mailto:#1}{\nolinkurl{#1}}}

\usepackage{datetime}
\date{\vspace{-2.5\baselineskip}}

\author[1]{Yasunori Akagi\footnote{Corresponding author. E-mail: \email{yasunori.akagi.cu@hco.ntt.co.jp}}}
\author[2]{Naoki Marumo}
\author[1]{Takeshi Kurashima}

\affil[1]{NTT Human Informatics Laboratories, Kanagawa, Japan}
\affil[2]{NTT Communication Science Laboratories, Kyoto, Japan}

\title{Analytically Tractable Models for Decision Making\\under Present Bias}

\begin{document}
\maketitle

\begin{abstract}
  Time-inconsistency is a characteristic of human behavior in which people plan for long-term benefits but take actions that differ from the plan due to conflicts with short-term benefits. Such time-inconsistent behavior is believed to be caused by present bias, a tendency to overestimate immediate rewards and underestimate future rewards. It is essential in behavioral economics to investigate the relationship between present bias and time-inconsistency.
  
  In this paper, we propose a model for analyzing agent behavior with present bias in tasks to make progress toward a goal over a specific period. Unlike previous models, the state sequence of the agent can be described analytically in our model. Based on this property, we analyze three crucial problems related to agents under present bias: task abandonment, optimal goal setting, and optimal reward scheduling. Extensive analysis reveals how present bias affects the condition under which task abandonment occurs and optimal intervention strategies. Our findings are meaningful for preventing task abandonment and intervening through incentives in the real world.
\end{abstract}

\section{Introduction}
\label{sec:intro}
People often do not achieve their goals because they change their plans in the middle of a task, even when nothing unexpected happens. For instance, some people may plan to stick to a diet for a month but end up indulging on the weekend. Similarly, a student may plan to do their assignments every day during vacation but procrastinate until the last day. Such behavior is known as \emph{time-inconsistency}, which is a topic of active research in behavioral economics.

Time-inconsistent behavior is often caused by \emph{present bias}, a tendency to overestimate the value of immediate rewards and underestimate the value of future rewards.
As an example, when planning a diet, the individual may not give much thought to treats on weekends, believing they can resist the temptation. 
However, when the weekend arrives, the desire for those treats becomes stronger, and the person may find it difficult to stick to their diet plan.

Researchers have been utilizing mathematical models to study the effects of present bias on human behavior, and recently \citet{kleinberg2014time} proposed a new model, which combines previous models.
The Kleinberg--Oren (KO) model is based on graph theory and simulates the time-inconsistent behavior of a person affected by present bias with an agent moving on a graph. 
The model represents a task with a directed acyclic graph whose vertices correspond to the agent's states.
Each edge has a cost for moving between vertices, and the goal vertex has a reward. 
On each vertex, the agent evaluates the value of each path to the goal and approaches the goal by following the path that seems most valuable to it.
The value of each path is evaluated with a particular discounting scheme called quasi-hyperbolic discounting \cite{laibson1997golden}, which introduces present bias into the agent.
The KO model is effective for reproducing typical time-inconsistencies in behavioral economics, such as procrastination, task abandonment, and choice reduction.
The model is also highly expressive and can represent a wide range of real-world tasks. 

However, the high flexibility of the model also makes it challenging to analyze the its properties. 
The agent's future behavior in the KO model cannot be derived in a closed form; simulation is necessary to determine the agent's behavior.
Additionally, it is difficult to identify optimal interventions for guiding the agent to the goal.
\citet{tang2017computational} and \citet{albers2019motivating} considered intervention by adding intermediate rewards, while \citet{albers2021value} increased the cost of edges.
It is shown in both settings that finding the optimal intervention is NP-hard, which demonstrates the computational difficulty of the KO model.

We introduce a model that is easy to analyze and compute while still being able to handle typical tasks in real life, based on the KO model.
To accomplish this, we limit tasks to increasing a real number, which we we refer to as \emph{progress}, over a specific period.
This type of task is commonly encountered in daily life; for example, completing a graduation thesis within six months (see the beginning of \cref{sec:proposed_model} for more examples).
Our model shares the same assumptions on agent behavior as that of the KO model but differs in three ways. 
\begin{itemize}
  \item The agent's state is represented as a pair of time index and progress rather than as a vertex of a graph.
  \item Because we regard progress as a real number rather than as an integer, the agent can take (uncountably) infinite states.
  \item The cost of an agent's action is expressed as $c(\Delta)$ with a function $c: \R \to \R \cup \set{+\infty}$, where $\Delta$ denotes the change in the progress between two adjacent time steps.
\end{itemize}
The main benefit of this model is that it simplifies the theoretical treatment of the model. 
We demonstrate that for a specific class of $c$, the trajectory of states taken by an agent can be described analytically.
This tractability is a significant advantage not found in existing models and allows us to perform various theoretical analyses, including finding optimal interventions.
Based on such analytical descriptions, we analyze three problems related to agents under present bias.

The first issue we examine \emph{task abandonment}, which refers to when a person starts a long-term project but gives up on it before completion, despite no change in the costs or rewards associated with the project. 
We theoretically analyze the conditions that lead to task abandonment in our model. 
The results show that a specific real number $\beta_0 \in (\frac{1}{e}, 1)$ is the essence ($e$ is Euler's number), which is determined by the shape of the cost function $c$ and the period.
Task abandonment can occur if a present-bias parameter $\beta$ is less than $\beta_0$, and never occurs otherwise.
We also analyze the asymptotic behavior of the threshold $\beta_0$ as the period's length grows, showing that $\beta_0$ converges to a value determined by the shape of the cost function $c$.
These results provide new insights into the conditions under which task abandonment occurs in real-world tasks.

The second problem is goal optimization; the goal has been fixed thus far, but from here, we control the goal to intervene in the agent's behavior.
We consider the problem of setting the goal to maximize the progress achieved by the agent, given a period and reward amount.
The crucial aspect in solving this problem is whether to allow exploitative rewards, i.e., rewards that influence the agent's behavior but are not claimed because the agent cannot satisfy the conditions for receiving the reward.
Our analysis reveals the structure of the optimal solution in both cases of allowing and not allowing exploitative rewards.
It shows that exploitative rewards can increase the progress of agents with strong present bias.
This result suggests that people with strong present bias are easily controlled and deceived by exploitative rewards.

The third problem is the optimization problem of reward scheduling, a more advanced intervention for the agent.
Given a total period and total reward budget, the goal is to determine the best way to distribute rewards over the period to maximize the final progress of the agent. 
Upon analyzing this problem, we found that the optimal strategy depends on the strength of present bias.
Our analysis shows that for agents with weak present bias that can adequately evaluate rewards and costs, it is ideal to offer rewards all at once.
In contrast, frequent intermediate rewards increase progress for agents with strong present bias.
This result indicates that we should vary the reward scheduling plan depending on the strength of the agents' present bias.

\section{Related Work}
The relationship between present bias and time-inconsistent behavior has been a central topic in behavioral economics for many years, both experimentally and theoretically \cite{frederick2002time, camerer2004behavioral, wilkinson2017introduction}.
In these studies, a discounting scheme for future values plays a crucial role because it determines time preferences, including present bias.
Classical economics used exponential discounting \cite{samuelson1937note}, which discounts value at a constant rate but can only lead to time-consistent behavior.
To resolve this issue, hyperbolic discounting, in which the discount rate decreases with time, was proposed and succeeded in explaining the time-inconsistent behavior of people \cite{ainslie1975specious}.
Quasi-hyperbolic discounting \cite{laibson1997golden, phelps1968second} was proposed to relax the analytical intractability of hyperbolic discounting and is widely used. 
This study also utilizes quasi-hyperbolic discounting to introduce present bias into the model. 

Numerous studies have analyzed human behavior on the basis of models with quasi-hyperbolic discounting.
These studies covered topics such as consumption-saving behavior \cite{laibson1997golden, laibson1998life}, addiction \cite{o1999addiction, gruber2001addiction}, and decisions in information acquisition \cite{carrillo2000strategic}. 
Particularly relevant to our study is the analysis of procrastination and task abandonment in long-term projects by \citet{o2008procrastination}.
This study is similar to ours in that it analyzes the influence of present bias on long-term goal-achieving behavior.
However, it differs from ours in that the model does not allow for an analytical description of the agent's state sequences, nor does it deal with the optimization problem of interventions.

As mentioned in \cref{sec:intro}, our model is inspired by the work of \citet{kleinberg2014time}. 
They investigated the graph-theoretic properties of the model, such as cost ratio, possible paths, and minimal motivating subgraphs.
However, they do not give the analytical description of agents' action sequences and optimization algorithms for interventions to guide agents. 
Although subsequent studies have tackled the optimization problems of various interventions \cite{albers2019motivating, albers2021value, tang2017computational}, they show the computational intractability of these problems and only present approximation algorithms. 
Our study succeeds in analytically describing the agent's behavior and finding optimal intervention strategies by restricting the types of tasks. 

The KO model has been extended in various directions to model real-world human behavior more accurately.
\citet{kleinberg2016planning} introduced a sophisticated agent in the model, which is an agent aware of its present bias and its influence (in contrast, the KO model assumes a naive agent unaware of its own present bias). 
\citet{kleinberg2017planning} introduced sunk-cost bias into the model, which is the tendency for people to continue investing in something they have already invested in, even when it no longer makes rational sense. 
\citet{gravin2016procrastination} proposed a model that draws the present-bias parameter from a fixed distribution in each round.
Our model does not adopt such extensions and is based on the original KO model.
Developing analytically tractable models that reflect these advanced factors will be future work. 

\section{Proposed Model}
\label{sec:proposed_model}
Our study deals with tasks that involve increasing a real number, called progress, to reach a goal over a specific period. 
We assume that the progress never decreases.
The following tasks fall into this category.
\begin{itemize}
\item Consider a person who sets a goal of exercising 30 hours in a month to improve his/her health condition. In this case, the specific period corresponds to one month, and the progress corresponds to the time spent exercising to date.
 \item Consider a salesperson trying to achieve a sales target of one million dollars in one year. In this case, the specific period corresponds to one year, and the progress corresponds to total sales so far. 
 \item Consider a student who sets a goal to complete his/her graduation thesis within six months.
 In this case, the specific period corresponds to six months, and the progress corresponds to the degree of completion of the graduation thesis.
\end{itemize}
We explain how to model the behavior of an agent dealing with such a task under present bias.

\subsection{Formulation}
Let $T \in \Z_{>0}$, $\theta \in \R_{\geq 0}$, and $R \in \R_{\geq 0}$ denote the period, goal, and 
reward, respectively.
Let a pair $(t, x)$ denote the state of an agent, where $t \in \set*{0, 1, \ldots, T}$ is the time index, and $x \in \R_{\geq 0}$ represents the progress that the agent has achieved at that time. 
The agent is initially in the state $(0, 0)$ and follows the sequence $(1, x_1), (2, x_2), \ldots, (T, x_T)$ as time passes.
Transitioning from state $(t, x_t)$ to state $(t+1, x_{t+1})$ involves a cost.
The cost is expressed as $c(x_{t+1} - x_{t})$ with a cost function $c: \R \to \R \cup \set{+\infty}$.

Let us explain how an agent decides which state to take next.
Suppose that the current state of the agent is $(t-1, x_{t-1})$ for $t \geq 1$. 
The agent evaluates the cost to follow the state sequence $(t, y_{t}), \ldots, (T, y_T)$ by 
\begin{align}
 \frac{1}{\beta} c(y_{t} - x_{t-1}) + \sum_{i=t+1}^T c(y_i - y_{i-1}) - R \cdot \bm{1}[y_T \geq \theta], 
 \label{eq:sequence cost}
\end{align}
where $\bm{1}[y_T \geq \theta] = 1$ if $y_T \geq \theta$, and $\bm{1}[y_T \geq \theta] = 0$ otherwise. 
The first term of \cref{eq:sequence cost} represents the transition cost from $(t-1, x_{t-1})$ to $(t, y_t)$, the second term represents the cost at subsequent times, and the third term represents the reward.

The first term is amplified by a coefficient of $\frac{1}{\beta}$ due to present bias, where $\beta \in (0, 1]$ is the \emph{present-bias parameter}; the agent overestimates the cost currently faced because of present bias.\footnote{
  Note that a large present-bias parameter $\beta$ means weak present bias.
  Although this may seem confusing, we adopt this notation in our paper for consistency with existing studies \cite{laibson1997golden, o2008procrastination, kleinberg2014time}.
}
This formulation of present bias is a type of what is called \emph{quasi-hyperbolic discounting} \cite{laibson1997golden} and is also used in the KO model. 
The third term of \cref{eq:sequence cost} means that the reward $R$ is obtained if the final progress $y_T$ is greater than or equal to $\theta$.
The reward has a negative sign because it has the opposite effect on the cost.

The agent computes the state sequence $(t, y_{t}^*), \ldots, (T, y_T^*)$ that minimizes the cost \cref{eq:sequence cost} and transitions from state $(t-1, x_{t-1})$ to $(t, y_{t}^*)$. 
Formally, the agent's state sequence is defined by 
$x_0 \coloneqq 0$ and
\begin{align}
 x_t
 &\coloneqq
 \argmin_{y_t \in \R}
 \min_{y_{t+1},\dots,y_T \in \R}
 \set*{
 \frac{1}{\beta} c(y_t - x_{t-1}) + \sum_{i=t+1}^T c(y_i - y_{i-1}) - R \cdot \bm{1}[y_T \geq \theta]
 }
 \label{eq:model_def_xt}
\end{align}
for $t = 1,\dots,T$.

Let us verify that this model is a variant of the KO model. 
A vertex of the graph in the KO model corresponds to a state $(t, x)$ in our model, and an edge of the graph corresponds to a transition between two states $(t, x_{t})$ and $(t+1, x_{t+1})$.
Under such correspondence, the rules of agent behavior are perfectly consistent between the two models. 
Note that our model differs from the KO model in that the possible state of the agent is a continuous quantity rather than a discrete quantity.

To make our model mathematically tractable, we assume that the cost function $c$ can be written as
\begin{align}
 \label{eq:cost function}
 c(\Delta)
 = 
 \begin{dcases*}
 \Delta^\alpha
 & if $\Delta \geq 0$,\\
 +\infty
 & otherwise, 
 \end{dcases*}
\end{align}
where $\alpha > 1$ is a parameter. 
Because we focus on tasks where the progress never decreases, we set $c(\Delta) = +\infty$ for $\Delta < 0$. 
Although this function may appear limited initially, it has the characteristics required for our desired modeling.

First, the cost function $c$ should be convex.
This is because tasks we deal with here are less labor-intensive if one does them steadily over a long time rather than all at once in a short time.
For example, it is less demanding to exercise one hour per day for ten days than to exercise for ten hours at once.
It is less burdensome to write three pages of a graduation thesis per day over ten days than to write 30 pages in one day.
The convexity of $c$ expresses this property. 
Second, the cost function $c$ should satisfy the property that $c(\Delta) = 0$ if and only if $\Delta = 0$.
In the accumulation-type tasks considered in this paper, it is reasonable to suppose that there is no progress without effort and vice versa.
Our cost function in \cref{eq:cost function} satisfies these two conditions and has $\alpha$ as a parameter, which can be adjusted to approximate the cost function for real-world tasks.

\subsection{Analytical solution}
Let 
\begin{align}
  p_t
  \coloneqq
  \frac{T-t}{T-t + \beta^{\frac{1}{\alpha - 1}}}
  \label{eq:def_pt}
\end{align}
to simplify the notation.
The following lemma gives a recursive formula for the state sequence $((t, x_t))_{t=0}^T$ taken by the agent.

\begin{lemma}
  \label{lemma:reccurence relation}
  The following holds for $t=1, 2, \ldots, T$:
  \begin{align} \label{eq:reccurence relation}
    x_t
    =
    \begin{dcases*}
      \theta + p_t (x_{t-1} - \theta)
      & if $ x_{t-1} \geq \theta - R^{\frac{1}{\alpha}} \prn[\big]{T - t + \beta^{\frac{1}{\alpha - 1}}}^{\frac{\alpha - 1}{\alpha}}$,\\
      x_{t-1}
      & otherwise.
    \end{dcases*}
    \label{eq:xi_recursion_quad_both}
  \end{align}
\end{lemma}
\begin{proof}
  For $t = T$, the desired result can be verified easily.
  We assume that $t < T$ below.

  Let us consider the $\min$-operation on \cref{eq:model_def_xt}.
  Assuming $y_T < \theta$, the minimum is achieved at $y_t = y_{t+1} = \dots = y_T$.
  Otherwise, the minimum is achieved when $y_T = \theta$.
  Therefore, the minimum can be evaluated as
  \begin{align}
    &\mathInd
    \min_{y_{t+1},\dots,y_T \in \R}
    \set[\bigg]{
      \frac{1}{\beta} c(y_t - x_{t-1}) + \sum_{i=t+1}^T c(y_i - y_{i-1}) - R \cdot \bm{1}[y_T \geq \theta]
    }\\
    &=
    \min \set[\bigg]{
      0,\ 
      \min_{y_{t+1},\dots,y_{T-1} \in \R}
      \set[\bigg]{
        \frac{1}{\beta} c(y_t - x_{t-1}) + \sum_{i=t+1}^{T-1} c(y_i - y_{i-1}) + c(\theta - y_{T-1}) - R
      }
    }\\
    &=
    \min \set*{
      0,\ 
      \frac{1}{\beta} c(y_t - x_{t-1})
      + (T - t) c \prn[\Big]{ \frac{\theta - y_t}{T - t} }
      - R
    },
    \label{eq:min_inner}
  \end{align}
  where we use Jensen's inequality for the last equality.
  Furthermore, we bound the second term as
  \begin{align}
    \frac{1}{\beta} c(y_t - x_{t-1})
    + (T - t) c \prn[\Big]{ \frac{\theta - y_t}{T - t} }
    &=
    \frac{1}{\beta} (y_t - x_{t-1})^\alpha
    + (T - t)^{1 - \alpha} \prn*{ \theta - y_t }^\alpha\\
    &\geq
    (\theta - x_{t-1})^\alpha
    (\beta^{\frac{1}{\alpha - 1}} + T - t)^{1 - \alpha},
  \end{align}
  where we have used \cref{eq:cost function} and H\"older's inequality:
  \begin{align}
    \theta - x_{t-1}
    &=
    (y_t - x_{t-1})
    + (\theta - y_t)
    \leq
    \prn*{
      \beta^{\frac{1}{\alpha - 1}} + (T - t)
    }^{\frac{\alpha - 1}{\alpha}}
    \prn[\Big]{
      \frac{1}{\beta} (y_t - x_{t-1})^\alpha
      + (T - t)^{1 - \alpha} \prn*{ \theta - y_t }^\alpha
    }^{\frac{1}{\alpha}}.
  \end{align}
  The equality holds when
  $
    \beta^{\frac{1}{\alpha - 1}}
    (T - t)^{1 - \alpha} \prn*{ \theta - y_t }^\alpha
    = 
    (T - t)
    \frac{1}{\beta} (y_t - x_{t-1})^\alpha
  $,
  or equivalently
  \begin{align}
    y_t
    =
    \theta + p_t (x_{t-1} - \theta).
    \label{eq:yt_optimal}
  \end{align}
  Hence, if $ (\theta - x_{t-1})^\alpha ( \beta^{\frac{1}{\alpha - 1}} + T - t )^{1 - \alpha} - R \leq 0$, the minimum on \cref{eq:model_def_xt} is achieved at \cref{eq:yt_optimal}.
  Otherwise, the minimum is achieved at $y_t = x_{t-1}$, which completes the proof.
\end{proof}

 Because $\theta - R^{\frac{1}{\alpha}} (T - t + \beta^{\frac{1}{\alpha - 1}})^{\frac{\alpha-1}{\alpha}}$ is increasing in $t$, \cref{lemma:reccurence relation} implies that there exists $t^* \in \set{0,\dots,T}$ such that
\begin{align}
x_t
=
\begin{dcases*}
  \theta + p_t (x_{t-1} - \theta)
  & if $t \leq t^*$,\\
  x_{t-1}
  & otherwise.
\end{dcases*}
\label{eq:xt_recursion_tast}
\end{align}
If $t^* = T$, the agent achieves the goal $\theta$ without giving up; otherwise, the agent gives up the goal at time $t^*$.
In particular, when $1 \leq t^* < T$, the agent abandons the task in the middle.
This result indicates that our model can reproduce task abandonment, in which an agent starts a task but gives up in the middle without any changes in underlying costs and rewards.

The following theorem characterizes the abandonment time $t^*$ and gives an analytical formula for the sequence $((t, x_t))_{t=1}^T$ taken by the agent.
\begin{theorem}
  \label{thm:analytical_solution}
  The abandonment time $t^*$ is the smallest $t \in \set{0, \dots, T-1}$ such that
  \begin{align}
    \prn*{T - t - 1 + \beta^{\frac{1}{\alpha - 1}}}^{1 - \alpha}
    \prod_{i=1}^t p_i^\alpha
    >
    \frac{R}{\theta^\alpha}
    \label{eq:condition_for_tast}
  \end{align}
  if there exists such $t$; otherwise, $t^* = T$.
  Moreover, the following holds for $t = 1,\dots,T$:
  \begin{align}
    x_t
    =
    \begin{dcases*}
      \theta \prn[\bigg]{1 - \prod_{i=1}^t p_i}
      & if $t \leq t^*$,\\
      \theta \prn[\bigg]{1 - \prod_{i=1}^{t^*} p_i}
      & otherwise.
    \end{dcases*}
    \label{eq:xt_explicit_formula}
  \end{align}
\end{theorem}
\begin{proof}
  The formula~\cref{eq:xt_explicit_formula} follows from \cref{eq:xt_recursion_tast} by induction on $t$.
  The $t^*$ is the smallest $t \in \set{0, \dots, T-1}$ such that $x_t < \theta - R^{\frac{1}{\alpha}} (T - t - 1 + \beta^{\frac{1}{\alpha - 1}})^{\frac{\alpha-1}{\alpha}}$ if there exists such $t$.
  A simple calculation shows with \cref{eq:xt_explicit_formula} show the equivalence between the condition on $t$ and \cref{eq:condition_for_tast}.
\end{proof}

\cref{thm:analytical_solution} states that the proposed model can describe the actual path of the agent analytically. 
This key feature is a major advantage of our model over the KO model.
Through such an analytical description, we will investigate the properties of the agent's behavior under present bias in the following sections.

\subsection{Discussion}
\begin{figure}
  \centering
  \subcaptionbox{$\alpha = 1.2$, $T = 10$}
  [0.42\linewidth]{\includegraphics[width=\linewidth]{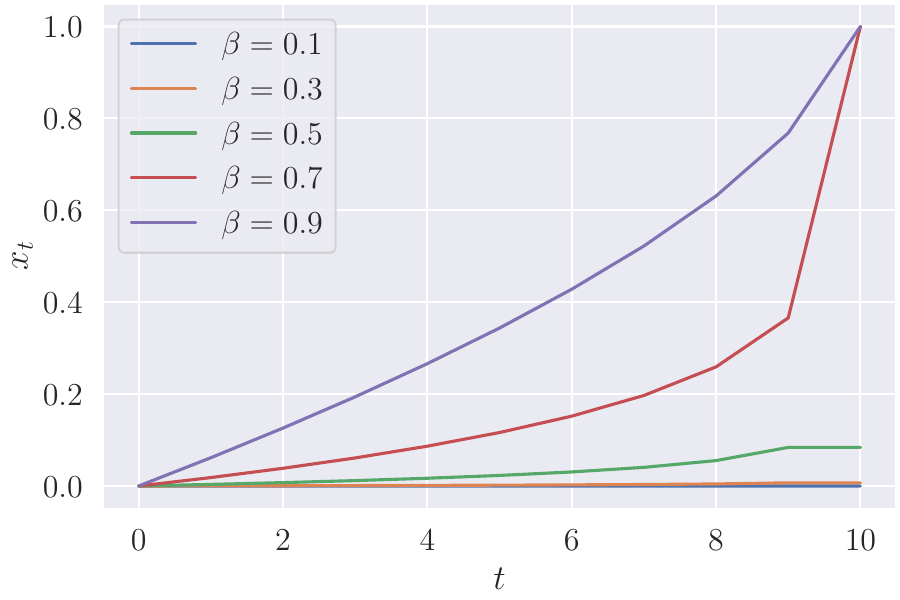}}\qquad%
  \subcaptionbox{$\alpha = 1.2$, $T = 1000$}
  [0.42\linewidth]{\includegraphics[width=\linewidth]{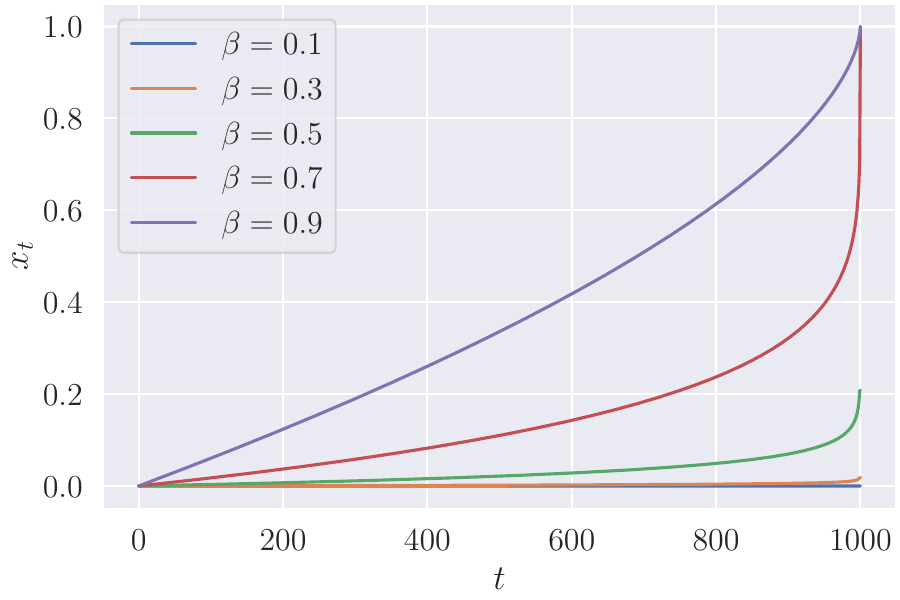}}\par\bigskip%
  \subcaptionbox{$\alpha = 2$, $T = 10$}
  [0.42\linewidth]{\includegraphics[width=\linewidth]{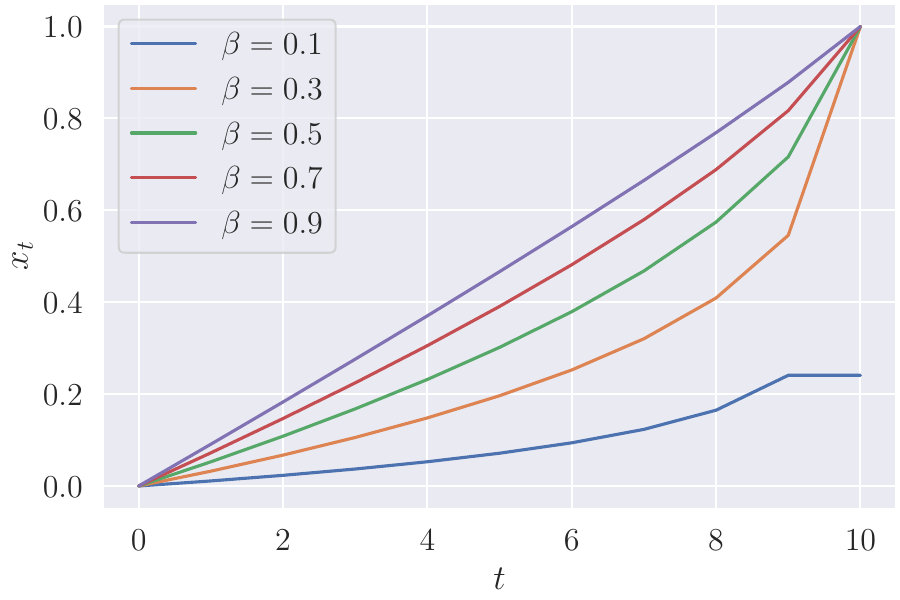}}\qquad%
  \subcaptionbox{$\alpha = 2$, $T = 1000$}
  [0.42\linewidth]{\includegraphics[width=\linewidth]{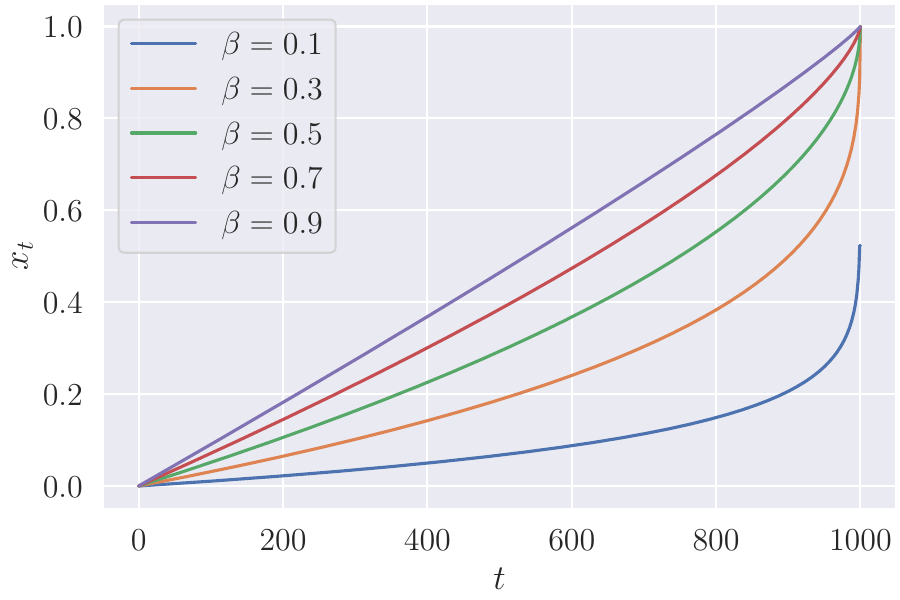}}\par\bigskip%
  \subcaptionbox{$\alpha = 10$, $T = 10$}
  [0.42\linewidth]{\includegraphics[width=\linewidth]{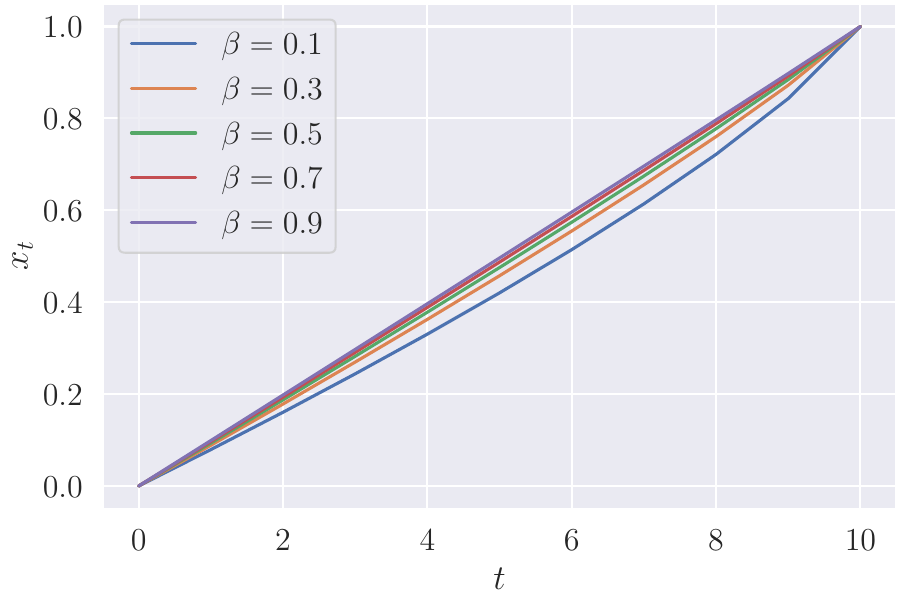}}\qquad%
  \subcaptionbox{$\alpha = 10$, $T = 1000$}
  [0.42\linewidth]{\includegraphics[width=\linewidth]{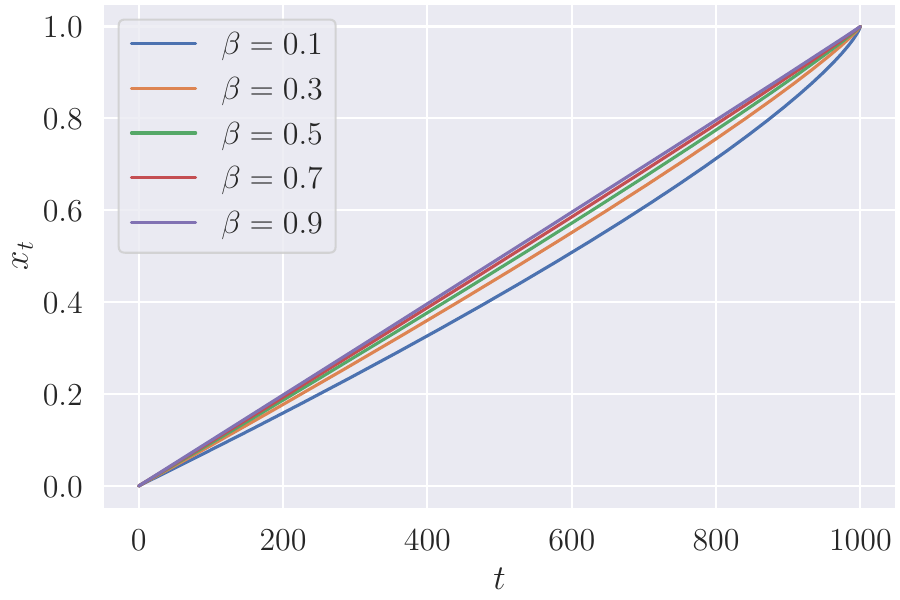}}\par%
  \caption{Agents' state sequences $((t, x_t))_{t=1}^T$ for $R = \theta = 1$.}
  \label{fig:xt}
\end{figure}

\Cref{fig:xt} shows the agents' state sequences $((t, x_t))_{t=1}^T$ computed by \cref{eq:xt_explicit_formula} for $R = \theta = 1$. 
These results suggest the following relationship between each parameter and the state sequences. 
\begin{itemize}
    \item
    When $\beta$ is small, progress is small in the beginning, and progress and significant at the end. This phenomenon can be interpreted as a case of procrastination induced by present bias.
    A small $\beta$ also leads to task abandonment, i.e., giving up the task without reaching the goal.
    In contrast, if $\beta$ is as large as 0.9, the agent makes (nearly) constant progress over the whole period. 
    \item
    The effect of $\alpha$ on progress is similar to that of $\beta$.
    The progress per time step monotonically increases with time when $\alpha$ is small, and the progress becomes constant when $\alpha$ is large.
    This occurs because making significant progress in one time step is easy when $\alpha$ is small, and the agent becomes overconfident about its future performance and tends to postpone making progress.
    \item
    The length $T$ of the period does not significantly affect the shape of the state sequences.
\end{itemize}

\section{Task abandonment and present bias}
\label{sec:abandonment}
In the previous section, we saw that our model reproduces task abandonment.
This section theoretically analyzes the relationship between task abandonment and the present-bias parameter $\beta$.

\subsection{Condition for task abandonment}
Let us fix the period $T$.
We introduce the concept of task-abandonment inducing for the present-bias parameter.
\begin{definition}
\label{def:tai}
  The present-bias parameter $\beta$ is said to be task-abandonment inducing (TAI) if the abandonment time $t^*$ in \cref{thm:analytical_solution} satisfies $0 < t^* < T$ for some $\theta, R \in \R_{\geq 0}$.
\end{definition}
An agent with a TAI $\beta$ may abandon the task in the middle depending on the reward $R$ or the goal  $\theta$.
On the other hand, an agent with non-TAI $\beta$ never abandons the task, i.e., either gives up the goal from the beginning or achieves the goal without giving up.
Investigating the TAI condition on $\beta$ helps us understand the relationship between the strength of present bias and the time-inconsistency of abandoning a task in the middle. 
The model's properties, including optimal intervention strategies, are greatly affected by whether or not $\beta$ is TAI, as we will see in \cref{sec:optimal goal setting,sec:optimal scheduling}. 

To simplify the notation, let $q_t$ denote the left-hand side on condition~\cref{eq:condition_for_tast}:
\begin{align}
  q_t
  \coloneqq
  \prn*{T - t - 1 + \beta^{\frac{1}{\alpha - 1}}}^{1 - \alpha}
  \prod_{i=1}^t p_i^\alpha.
  \label{eq:def_qt}
\end{align}
\cref{thm:analytical_solution} implies that $\beta$ is not TAI if and only if $\max_{0 \leq t < T} q_t = q_0$.
To check if $\beta$ is TAI, let us observe the properties of $q_t$.
\begin{lemma}
  \label{lem:q_beta}
  The following hold:
  \begin{enuminlem}
    \item
    \label{lem:q_beta_small_beta}
    if $\beta \leq (1 - \frac{1}{\alpha})^{\alpha - 1}$, then $q_0 < q_1 < \dots < q_{T-1}$,
    \item
    \label{lem:q_beta_medium_beta}
    if $(1 - \frac{1}{\alpha})^{\alpha - 1} < \beta < (1 - \frac{1}{\alpha})^{\frac{\alpha - 1}{2}}$, then there exists $t \in \set{0,\dots,T-1}$ such that $q_0 \geq q_1 \geq \dots \geq q_t < q_{t+1} < \dots < q_{T-1}$,
    \item
    \label{lem:q_beta_large_beta}
    if $\beta \geq (1 - \frac{1}{\alpha})^{\frac{\alpha - 1}{2}}$, then $q_0 > q_1 > \dots > q_{T-1}$.
  \end{enuminlem}
\end{lemma}
\begin{proof}
  Let $\gamma \coloneqq \beta^{\frac{1}{\alpha - 1}}$.
  From definitions \cref{eq:def_pt,eq:def_qt} of $p_t$ and $q_t$, we have
  \begin{align}
    \log \prn*{\frac{q_{t-1}}{q_t}}
    &=
    (1 - \alpha)
    \log
    \prn*{
      \frac{T - t + \gamma}{T - t - 1 + \gamma}
    }
    - \alpha \log 
    \prn*{
      \frac{T - t}{T - t + \gamma}
    }\\
    &=
    \log \prn*{1 + \frac{\gamma}{T - t}}
    + (\alpha - 1) \log \prn*{1 - \frac{1 - \gamma}{T - t}}
    =
    f(T - t),
  \end{align}
  where
  \begin{align}
    f(x)
    \coloneqq
    \log \prn*{1 + \frac{\gamma}{x}}
    + (\alpha - 1) \log \prn*{1 - \frac{1 - \gamma}{x}}.
  \end{align}
  Let us investigate the function $f$.
  We have
  \begin{align}
    \lim_{x \searrow 1 - \gamma} f(x) = - \infty,\quad
    \lim_{x \to +\infty} f(x) = 0,\quad
    f'(x)
    =
    \frac{\alpha \gamma (1 - \gamma) + (\alpha (1 - \gamma) - 1) x}{x (x + \gamma) (x + \gamma - 1)}.
  \end{align}
  
  \paragraph{Case (a): $\alpha (1 - \gamma) \geq 1$, or equivalently $\beta \leq (1 - \frac{1}{\alpha})^{\alpha - 1}$.}
  The function $f(x)$ is increasing for $x > 1 - \gamma$.
  Hence, $f(x) < 0$ for $x \geq 1$, which yields \cref{lem:q_beta_small_beta}.

  \paragraph{Case (b): $\alpha (1 - \gamma) < 1 < \alpha (1 - \gamma^2)$, or equivalently $(1 - \frac{1}{\alpha})^{\alpha - 1} < \beta < (1 - \frac{1}{\alpha})^{\frac{\alpha - 1}{2}}$.}
  The function $f(x)$ is increasing for $1 - \gamma < x < \frac{\alpha \gamma (1 - \gamma)}{1 - \alpha (1 - \gamma)}$ and is decreasing for $x > \frac{\alpha \gamma (1 - \gamma)}{1 - \alpha (1 - \gamma)}$.
  Hence, there exists $a > 1 - \gamma$ such that $f(x) < 0$ for $x < a$ and $f(x) \geq 0$ for $x \geq a$, which yields \cref{lem:q_beta_medium_beta}.

  \paragraph{Case (c): $\alpha (1 - \gamma^2) \leq 1$, or equivalently $\beta \geq (1 - \frac{1}{\alpha})^{\frac{\alpha - 1}{2}}$.}
  A simple calculation shows that $\alpha (1 - \gamma^2) \leq 1$ is equivalent to $\frac{\alpha \gamma (1 - \gamma)}{1 - \alpha (1 - \gamma)} \leq 1$, and thus $f(x)$ is decreasing for $x \geq 1$.
  Hence, $f(x) > 0$ for $x \geq 1$, which yields \cref{lem:q_beta_large_beta}.
\end{proof}

A closer look at the case of $(1 - \frac{1}{\alpha})^{\alpha - 1} < \beta < (1 - \frac{1}{\alpha})^{\frac{\alpha - 1}{2}}$ yields the following lemma.
\begin{lemma}
  \label{lem:q_beta_closer}
  There exists $(1 - \frac{1}{\alpha})^{\alpha - 1} < \beta_0 < (1 - \frac{1}{\alpha})^{\frac{\alpha - 1}{2}}$ that depends only on $T$ and $\alpha$ and satisfies the following:
  \begin{enuminlem}
    \item
    if $(1 - \frac{1}{\alpha})^{\alpha - 1} < \beta < \beta_0$, then $q_0 < q_{T-1}$,
    \item
    if $\beta = \beta_0$, then $q_0 = q_{T-1}$,
    \item
    if $\beta_0 < \beta < (1 - \frac{1}{\alpha})^{\frac{\alpha - 1}{2}}$, then $q_0 > q_{T-1}$.
  \end{enuminlem}
\end{lemma}
\begin{proof}
  Let $\gamma \coloneqq \beta^{\frac{1}{\alpha - 1}}$.
  We have
  \begin{align}
    \frac{q_{T-1}}{q_0}
    = 
    \prn*{\frac{T - 1 + \gamma}{\gamma}}^{\alpha - 1}
    \prod_{t=1}^{T-1} p_t^\alpha
    =
    \prn*{ 1 + \frac{T - 1}{\gamma} }^{\alpha - 1}
    \prod_{t=1}^{T-1} \prn*{\frac{T - t}{T - t + \gamma}}^\alpha,
    \label{eq:qTq1_ratio}
  \end{align}
  and the value is decreasing in $\gamma$, or equivalently in $\beta$.
  We also know from \cref{lem:q_beta} that $\frac{q_{T-1}}{q_0} > 1$ for $\beta = (1 - \frac{1}{\alpha})^{\alpha - 1}$ and $\frac{q_{T-1}}{q_0} < 1$ for $\beta = (1 - \frac{1}{\alpha})^{\frac{\alpha - 1}{2}}$, which completes the proof.
\end{proof}

In the remaining part of the paper, let $\beta_0$ denote the $\beta_0$ in \cref{lem:q_beta_closer}.
Since $\beta$ is not TAI if and only if $\max_{0 \leq t < T} q_t = q_0$, \cref{lem:q_beta,lem:q_beta_closer} yield the following theorem.
\begin{theorem}
  \label{thm:beta_TAI}
  The present-bias parameter $\beta$ is TAI if and only if $\beta < \beta_0$.
\end{theorem}
\Cref{thm:beta_TAI} shows that if $\beta$ is large, i.e., the present bias is weak, the agent will accomplish the goal if it has decided to achieve the goal at $t=0$.
If $\beta$ is small, the agent may abandon the task in the middle. 
These results are consistent with the past findings on present bias and time-inconsistency \cite{kleinberg2014time}.
The threshold $\beta_0$ will be crucial in the following sections. 

As in \cref{lem:q_beta_closer}, $\beta_0$ is lower-bounded as $\beta_0 > (1 - \frac{1}{\alpha})^{\alpha - 1}$.
At the same time, elementary calculus shows that $(1 - \frac{1}{\alpha})^{\alpha - 1} > \frac{1}{e}$ for all $\alpha > 1$ and $\lim_{\alpha \to \infty} (1 - \frac{1}{\alpha})^{\alpha - 1} = \frac{1}{e}$, where $e$ is Euler's number.
Therefore, a small $\beta \leq \frac{1}{e} \approx 0.368$ is consistently TAI in our model; interestingly, the threshold involves Euler's number. 

\subsection{Asymptotic formula for $\beta_0$}
The following theorem gives an asymptotic formula for $\beta_0$ when $T \to \infty$.
\begin{theorem}
\label{thm:asymptotic_beta_0}
  The following holds:
  \begin{align}
    \beta_0
    &=
    \prn*{ 1 - \frac{1}{\alpha} }^{\alpha - 1}
    \prn*{
      1
      + \frac{
        \alpha \log \Gamma(1 - \frac{1}{\alpha})
        + \log (1 - \frac{1}{\alpha})
      }{\log T}
    }
    + o\prn*{\frac{1}{\log T}},
    \label{eq:beta0_formula_general}
  \end{align}
  where $\Gamma$ is the gamma function.
  In particular, if $\alpha = 2$, then 
  \begin{align}
    \beta_0
    &=
    \frac{1}{2}
    + \frac{\log (\pi / 2)}{2 \log T}
    + o \prn*{\frac{1}{\log T}}.
  \end{align}
\end{theorem}
\begin{proof}
  Let $\gamma \coloneqq \beta^{\frac{1}{\alpha - 1}}$ and $\gamma_0 \coloneqq \beta_0^{\frac{1}{\alpha - 1}}$.
  The $\gamma_0$ is $\gamma$ such that $\frac{q_{T-1}}{q_0} = 1$.
  From \cref{eq:qTq1_ratio}, we have
  \begin{align}
    1
    =
    \prn*{ 1 + \frac{T - 1}{\gamma_0} }^{\alpha - 1}
    \prod_{t=1}^{T-1} \prn*{\frac{T - t}{T - t + \gamma_0}}^\alpha
    &=
    \prn*{ 1 + \frac{T - 1}{\gamma_0} }^{\alpha - 1}
    \prn*{
      \frac{\Gamma(T) \Gamma(1 + \gamma_0)}{\Gamma(T + \gamma_0)}
    }^\alpha.
  \end{align}
  The asymptotic formula $\frac{\Gamma(T)}{\Gamma(T + \gamma)} = T^{- \gamma} (1 + o(1))$ (e.g., \citep[Eq.~(1)]{tricomi1951asymptotic}) leads to
  \begin{align}
    1
    &=
    \prn*{ 1 + \frac{T - 1}{\gamma_0} }^{\alpha - 1}
    \frac{\Gamma(1 + \gamma_0)^\alpha}{T^{\alpha \gamma_0}}
    (1 + o(1))^\alpha
    =
    T^{\alpha (1 - \gamma_0) - 1}
    \gamma_0 \Gamma(\gamma_0)^\alpha
    (1 + o(1)),
  \end{align}
  where we have used $\Gamma(1 + \gamma_0) = \gamma_0 \Gamma(\gamma_0)$.
  Comparing the order of $T$ on both sides gives $\lim_{T \to \infty} \gamma_0 = 1 - \frac{1}{\alpha}$.
  Furthermore, taking logarithms of the above equation yields
  \begin{alignat}{2}
    0
    &=
    (\alpha (1 - \gamma_0) - 1) \log T
    + \log \prn*{ \gamma_0 \Gamma(\gamma_0)^\alpha }
    + \log \prn*{1 + o(1)}\\
    &=
    (\alpha (1 - \gamma_0) - 1) \log T
    + \log \prn*{ \prn[\Big]{1 - \frac{1}{\alpha}} \Gamma\prn[\Big]{1 - \frac{1}{\alpha}}^\alpha }
    + o(1).
  \end{alignat}
  Rearranging the terms leads to
  \begin{align}
    \gamma_0
    =
    1 - \frac{1}{\alpha}
    + \frac{
      \log \Gamma(1 - \frac{1}{\alpha})
      + \frac{1}{\alpha} \log (1 - \frac{1}{\alpha})
    }{\log T}
    + o\prn*{\frac{1}{\log T}}.
  \end{align}
  Using $\beta_0 = \gamma_0^{\alpha - 1}$ and the Taylor series $x^{\alpha - 1} = b^{\alpha - 1} (1 + \frac{\alpha - 1}{b} (x - b)) + O((x - b)^2)$ yields the general formula~\cref{eq:beta0_formula_general}.
  The formula for $\alpha = 2$ follows from $\Gamma(1/2) = \sqrt{\pi}$.
\end{proof}

\cref{thm:asymptotic_beta_0} gives several insights about the threshold $\beta_0$.
\begin{itemize}
  \item
  The threshold $\beta_0$ converges to $\prn*{ 1 - 1/\alpha}^{\alpha - 1}$ as $T \to \infty$.
  The value $\prn*{ 1 - 1/\alpha}^{\alpha - 1}$ is equal to the lower bound on $\beta_0$ given in \cref{lem:q_beta_closer}.
  \item
  The convergence speed is slow as $\beta_0 - \prn*{ 1 - 1/\alpha}^{\alpha - 1} = \Theta(1 / \log T)$.
  \item
  When $\alpha = 2$, the asymptotic formula for $\beta_0$ involves $\pi$, interestingly.
\end{itemize}

\section{Optimal Goal Setting}
\label{sec:optimal goal setting}
In this section, we consider the problem of setting a goal $\theta$ to maximize the final progress $x_T$:
\begin{align}
  \max_{\theta \geq 0}\ x_T,
  \label{eq:problem_goal_setting}
\end{align}
given the period $T$ and reward $R$.

Note that the optimal solution to this problem varies depending on whether \emph{exploitative rewards} are allowed.
Exploitative rewards are rewards that are placed in order to motivate the agent but can never be claimed because the agent never reaches the target.
Under the influence of present bias, it is possible that $x_T$ can be increased by installing the exploitative reward as bait, even if it cannot actually be obtained.
However, using exploitative rewards can raise ethical concerns and decrease human motivation in real-world scenarios, so they should be used with caution. 
In this section, we will examine both scenarios where exploitative rewards are allowed and not allowed.

\subsection{Non-TAI $\beta$}
Because $\beta$ is not TAI, no reward setting can be exploitative.
In other words, the optimal solution remains the same whether or not exploitative rewards are allowed.
The following theorem provides an explicit formula for the optimal solution.

\begin{theorem}
  \label{thm:goal_setting1}
  Regardless of whether or not exploitative rewards are allowed, the optimal solution to problem~\cref{eq:problem_goal_setting} is
  \begin{align}
    \theta 
    =
    R^{\frac{1}{\alpha}}
    \prn*{T - 1 + \beta^{\frac{1}{\alpha - 1}}}^{\frac{\alpha - 1}{\alpha}}.
    \label{eq:optimal_theta_formula1}
    \end{align}
  The optimal value is the same as $\theta$.
\end{theorem}
\begin{proof}
  Remember that $\max_{0 \leq t < T} q_t = q_0$, as mentioned in \cref{sec:abandonment}.
  From \cref{eq:condition_for_tast}, if $q_0 > \frac{R}{\theta^\alpha}$, then $t^* = 0$, and therefore $x_T = 0$; otherwise, we have $t^* = T$.
  Thus, the optimal solution to problem~\cref{eq:problem_goal_setting} is the largest $\theta$ such that $q_0 \leq \frac{R}{\theta^\alpha}$, i.e., $\theta = (R / q_0)^{\frac{1}{\alpha}} = R^{\frac{1}{\alpha}} (T - 1 + \beta^{\frac{1}{\alpha - 1}})^{\frac{\alpha - 1}{\alpha}}$.
\end{proof}

\subsection{TAI $\beta$}
Because $\beta_0$ is TAI, the optimal solution can vary depending on whether or not exploitative rewards are allowed.
The following theorem gives an explicit formula for when exploitative rewards are not allowed.
For cases where they are allowed, we can reduce the continuous optimization problem~\cref{eq:problem_goal_setting} to a discrete one, though the optimal solution is difficult to express in a closed form.
\begin{theorem}
  \label{thm:goal_setting2}
  \leavevmode
  \begin{enuminthm}
    \item
    \label{thm:goal_setting2_nonexploitative}
    Suppose that exploitative rewards are not allowed.
    Then the optimal solution to problem~\cref{eq:problem_goal_setting} is
    \begin{align}
      \theta
      = \prn*{\frac{R}{q_{T-1}}}^{\frac{1}{\alpha}}
      =
      (\beta R)^{\frac{1}{\alpha}}
      \frac{\Gamma(T + \beta^{\frac{1}{\alpha - 1}})}{\Gamma(T) \Gamma(1 + \beta^{\frac{1}{\alpha - 1}})}.
      \label{eq:optimal_theta_formula2}
    \end{align}
    The optimal value is the same as $\theta$.
    \item
    \label{thm:goal_setting2_exploitative}
    Suppose that exploitative rewards are allowed.
    Problem~\cref{eq:problem_goal_setting} is reduced to
    \begin{align}
      \max_{t \in \set{\tilde{t},\dots,T}} \, u_t, 
      \label{eq:problem_goal_setting_discrete}
    \end{align}
    where 
    \begin{align}
      u_t
      \coloneqq
      \prn*{\frac{R}{\max \set*{q_0, q_{t-1}}}}^{\frac{1}{\alpha}} \prn*{1 - \prod_{i=1}^t p_i}. 
      \label{eq:def_ut}
    \end{align}
    and $\tilde{t}$ is the smallest $t$ which satisfies $q_t > q_0$. 
    For the optimal solution $t$ to problem~\cref{eq:problem_goal_setting_discrete}, the optimal solution to \cref{eq:problem_goal_setting} is written by
    \begin{align}
      \theta
      = 
      \prn*{\frac{R}{\max \set*{q_0, q_{t-1}}}}^{\frac{1}{\alpha}}
      \label{eq:optimal_solution_discrete_to_continuous}
    \end{align}
  \end{enuminthm}
\end{theorem}
\begin{proof}[Proof of \cref{thm:goal_setting2_nonexploitative}]
  The agent does not abandon the reward if and only if $q_{T-1} \leq \frac{R}{\theta^\alpha}$, since $\max_{0 \leq t < T} q_t = q_{T-1}$ as discussed in \cref{sec:abandonment}.
  Therefore, the optimal solution to problem~\cref{eq:problem_goal_setting} is the largest $\theta$ such that $q_{T-1} \leq \frac{R}{\theta^\alpha}$, i.e., $\theta = (R / q_{T-1})^{\frac{1}{\alpha}}$.
  A simple calculation with definitions \cref{eq:def_pt,eq:def_qt} of $p_t$ and $q_t$ leads to
  \begin{align}
    \theta
    =
    (R / q_{T-1})^{\frac{1}{\alpha}}
    =
    (\beta R)^{\frac{1}{\alpha}}
    \prod_{t=1}^{T-1} \frac{T-t + \beta^{\frac{1}{\alpha - 1}}}{T-t}
    =
    (\beta R)^{\frac{1}{\alpha}}
    \frac{\Gamma(T + \beta^{\frac{1}{\alpha - 1}})}{\Gamma(T) \Gamma(1 + \beta^{\frac{1}{\alpha - 1}})}.
  \end{align}
\end{proof}

\begin{proof}[Proof of \cref{thm:goal_setting2_exploitative}]
  We classify cases based on the time of task abandonment, $t \ (=0, 1, \ldots, T)$.

\paragraph{Case (a): $t = 0$.} In this case,the achieved progress $x_T$ is 0.

\paragraph{Case (b): $t = T$. }
Because this case is the same as the case where exploitative rewards are prohibited, the optimal $\theta$ is $\prn*{\frac{R}{q_{T-1}}}^{\frac{1}{\alpha}}$, and the achieved progress $x_T$ equals to $\theta$.  

\paragraph{Case (c): $0 < t < T$. } The condition for giving up at time $t$ is not giving up at times $0, \ldots, t-1$ and giving up at time $t$, i.e.,
\begin{align}
    q_s &\leq \frac{R}{\theta^{\alpha}} \quad (s=0, \ldots, t-1), \label{eq:cond1}\\
    q_t &> \frac{R}{\theta^{\alpha}}. \label{eq:cond2}
\end{align}

If $q_t \leq q_0$, then from \cref{eq:cond1}, $q_0 \leq \frac{R}{\theta^{\alpha}}$, implying $q_t \leq \frac{R}{\theta^{\alpha}}$, which contradicts \cref{eq:cond2}. Therefore, it is necessary that $q_t > q_0$.
Conversely, if $q_t > q_0$, by setting $\theta$ to satisfy 
\begin{align}
\prn*{\frac{R}{q_t}}^{\frac{1}{\alpha}} < \theta \leq \prn*{\frac{R}{\max_{s=0, 1, \ldots, t-1} q_s}}^{\frac{1}{\alpha}}, 
\end{align}
the agent will give up at time $t$.

When we fix the abandonment time $t$, the achieved progress is $\theta \prn*{1 - \prod_{i=1}^t p_i}$, which is monotonically increasing with respect to $\theta$. Thus, the optimal $\theta$ is given by
\begin{align}
    \theta = \prn*{\frac{R}{\max_{s=0, 1, \ldots, t-1} q_s}}^{\frac{1}{\alpha}} = \prn*{\frac{R}{\max \set*{q_0, q_{t-1}}}}^{\frac{1}{\alpha}},
\end{align}
and the maximum achievable progress is
\begin{align}
    \prn*{\frac{R}{\max \set*{q_0, q_{t-1}}}}^{\frac{1}{\alpha}} \prn*{1 - \prod_{i=1}^t p_i}.
\end{align}

Summarizing (a), (b), and (c), Problem (12) reduces to
\begin{align}
    \max_{t \in \set*{\tilde{t}, \ldots, T}} \prn*{\frac{R}{\max \set*{q_0, q_{t-1}}}}^{\frac{1}{\alpha}} \prn*{1 - \prod_{i=1}^t p_i},
\end{align}
, where $\tilde{t}$ is the smallest $t$ such that $q_t > q_0$.
Using $t$ that achieves the maximum, the optimal $\theta$ is expressed as
\begin{align}
    \prn*{\frac{R}{\max \set*{q_0, q_{t-1}}}}^{\frac{1}{\alpha}}.
\end{align}
\end{proof}

As we can compute $u_1,\dots,u_T$ in \cref{eq:problem_goal_setting_discrete} in $O(T)$ time, the optimal solution to problem~\cref{eq:problem_goal_setting} is obtained in $O(T)$ time from \cref{thm:goal_setting2_exploitative} when exploitative rewards are allowed.

\subsection{Discussion}

\begin{figure}[t]
  \centering
  \subcaptionbox{$\alpha = 1.2$, $T = 10$}
  [0.33\linewidth]{\includegraphics[width=\linewidth]{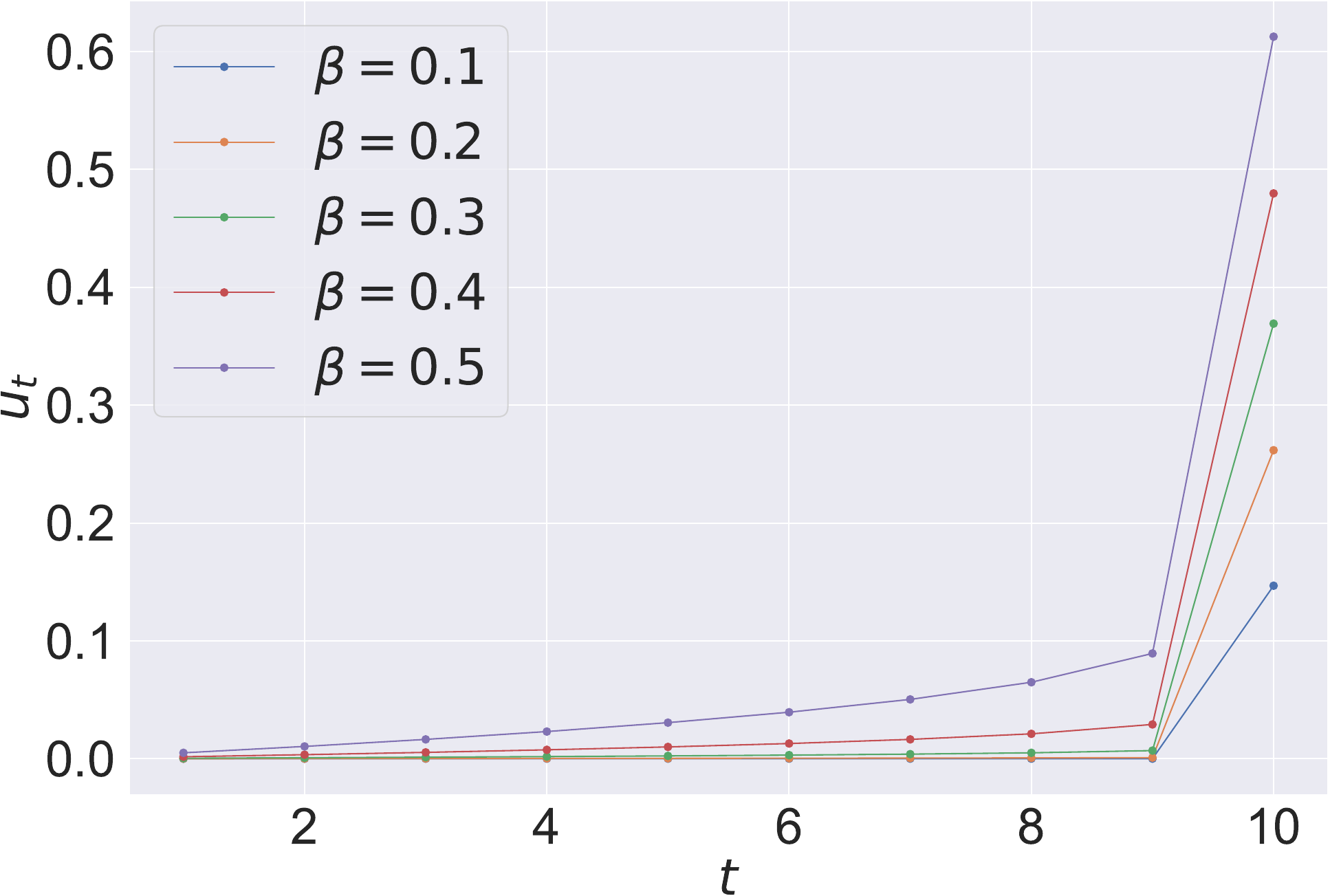}}\hfill%
  \subcaptionbox{$\alpha = 1.2$, $T = 100$}
  [0.33\linewidth]{\includegraphics[width=\linewidth]{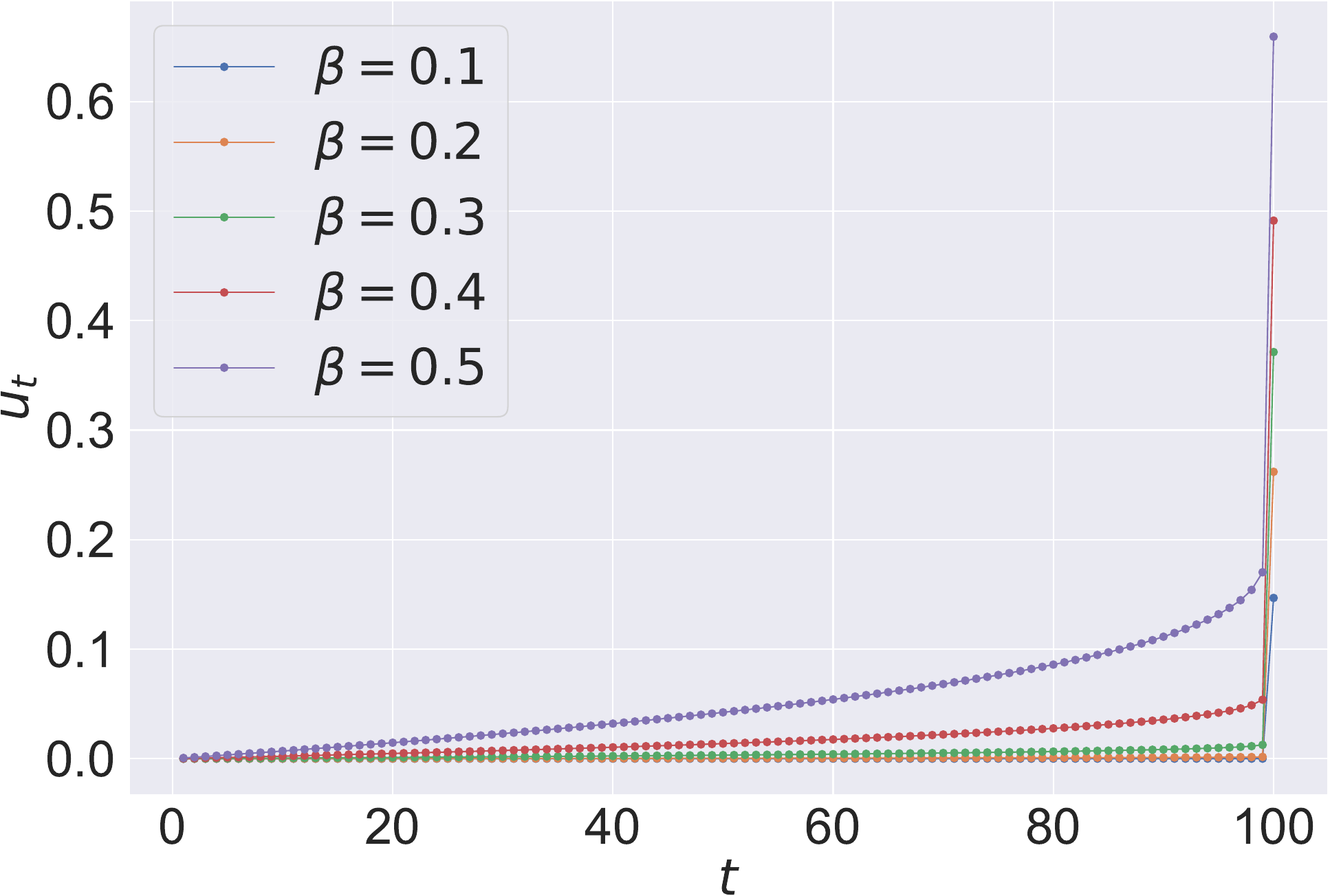}}\hfill%
  \subcaptionbox{$\alpha = 1.2$, $T = 1000$}
  [0.33\linewidth]{\includegraphics[width=\linewidth]{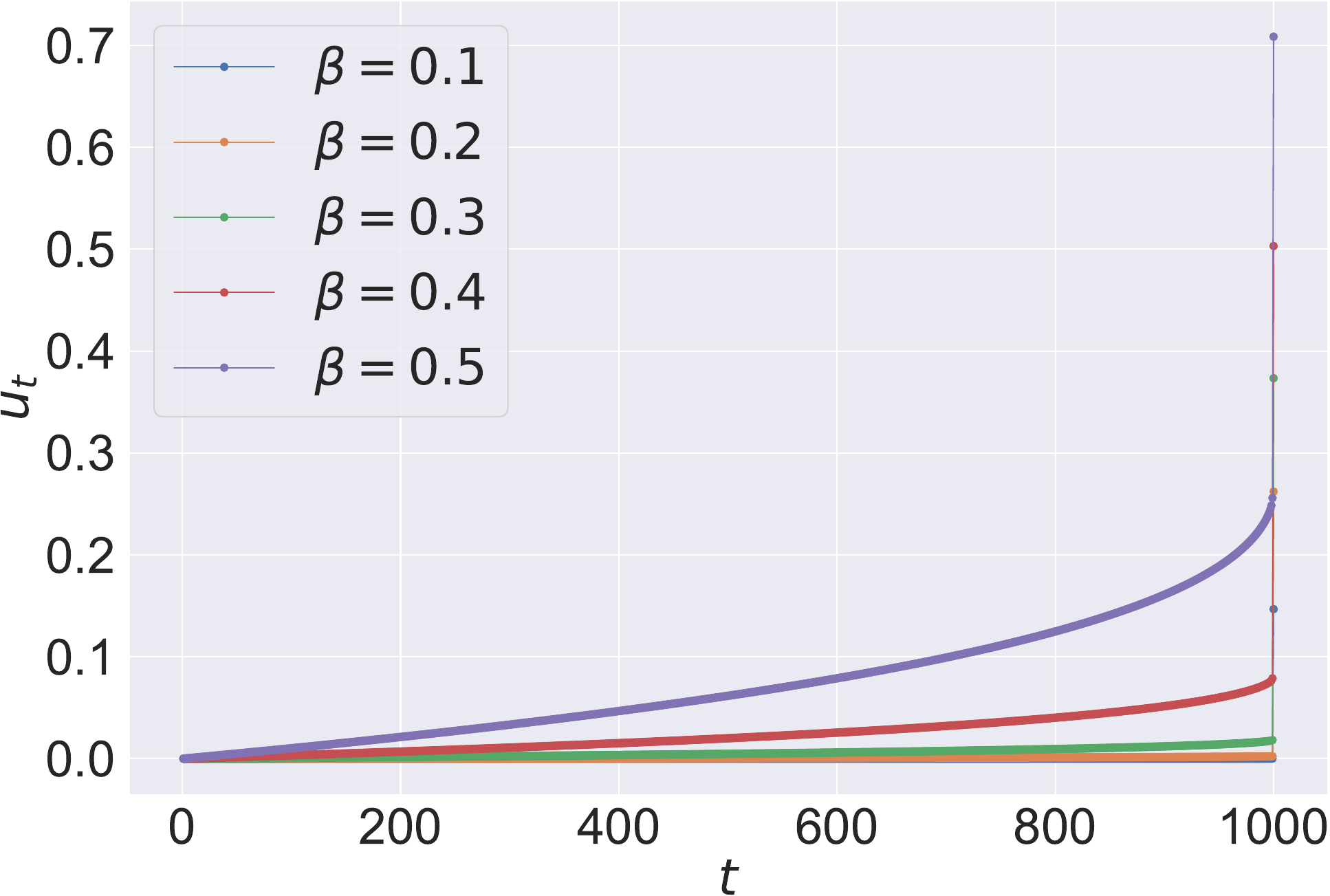}}\par\bigskip%
  \subcaptionbox{$\alpha = 2$, $T = 10$}
  [0.33\linewidth]{\includegraphics[width=\linewidth]{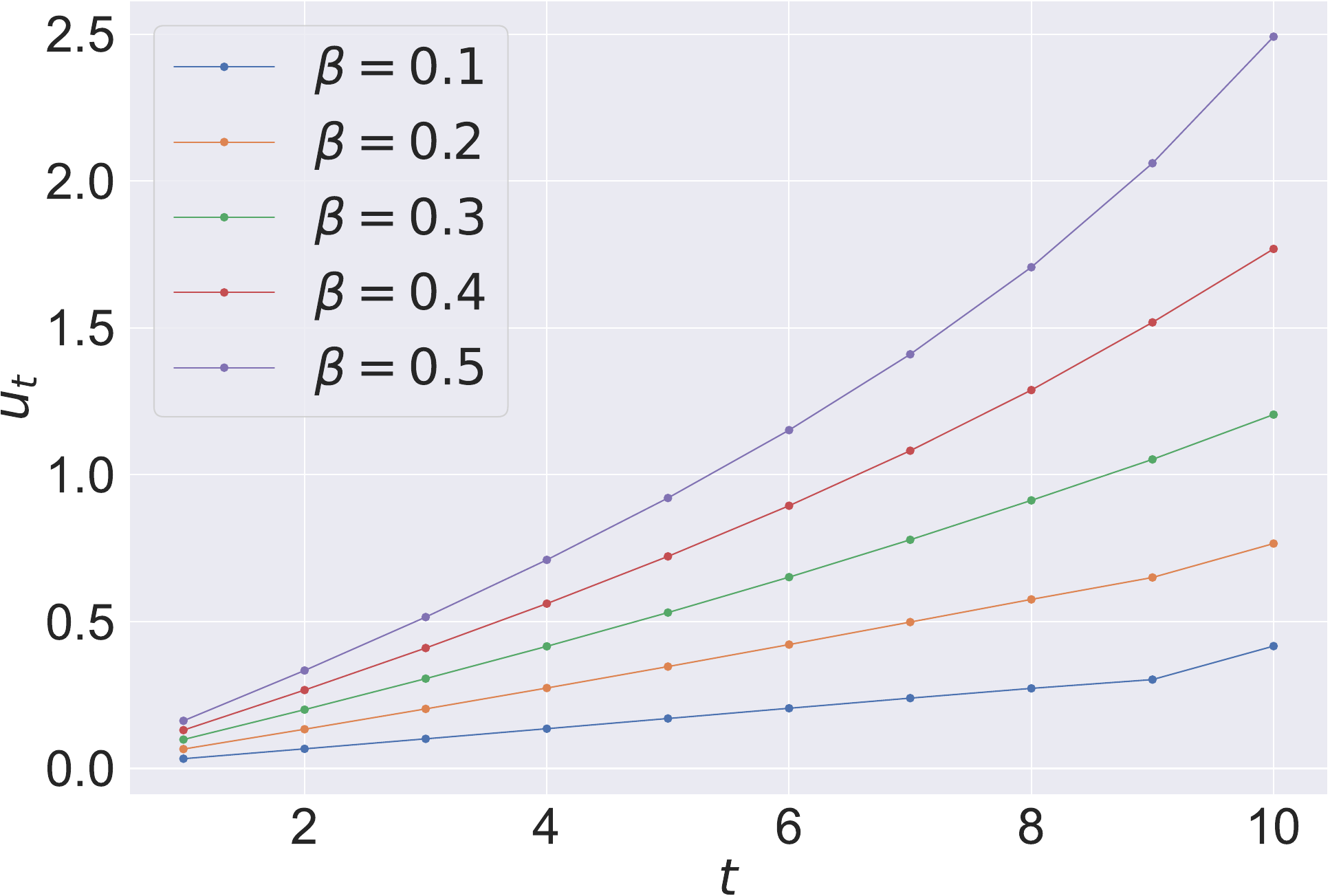}}\hfill%
  \subcaptionbox{$\alpha = 2$, $T = 100$}
  [0.33\linewidth]{\includegraphics[width=\linewidth]{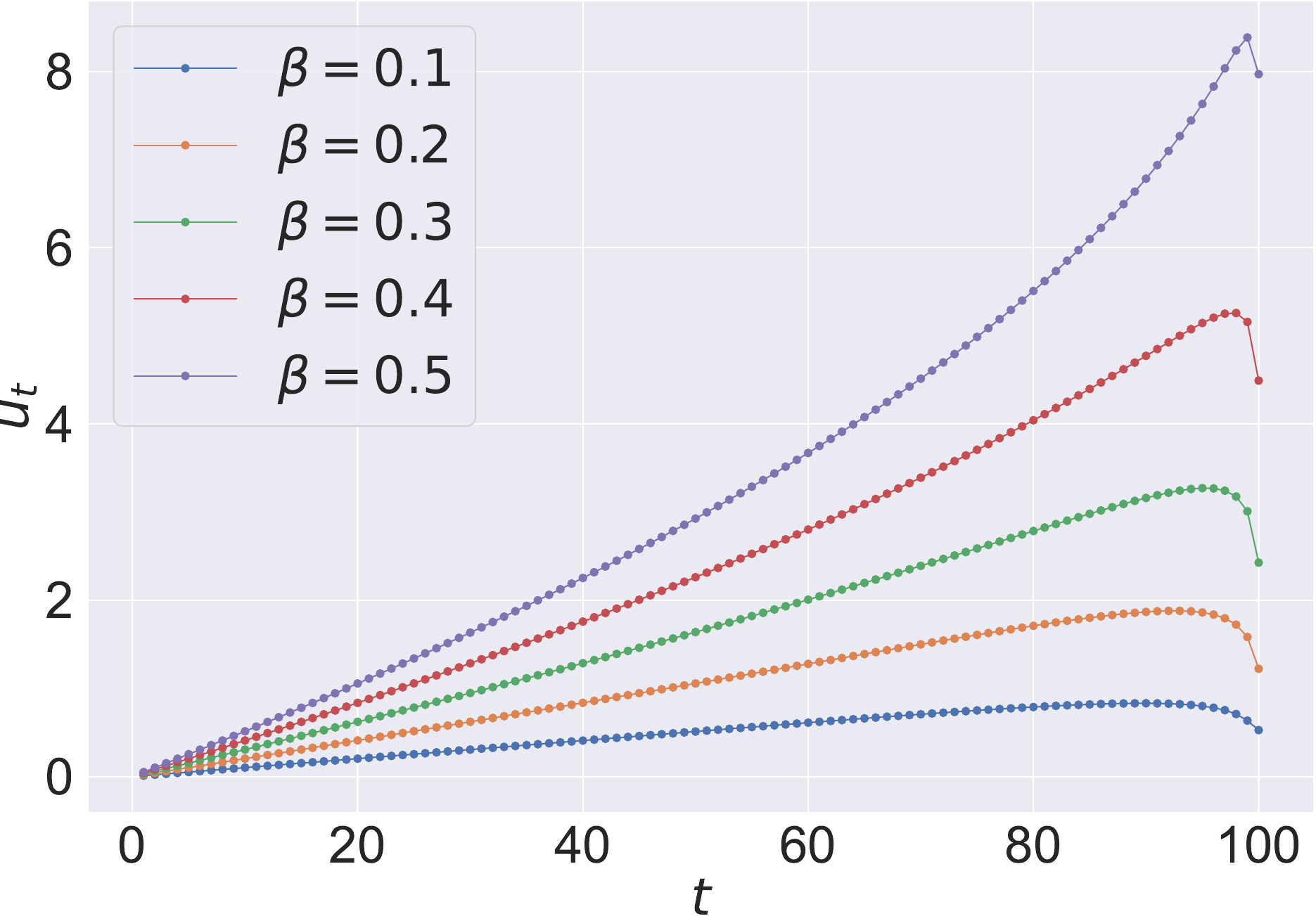}}\hfill%
  \subcaptionbox{$\alpha = 2$, $T = 1000$}
  [0.33\linewidth]{\includegraphics[width=\linewidth]{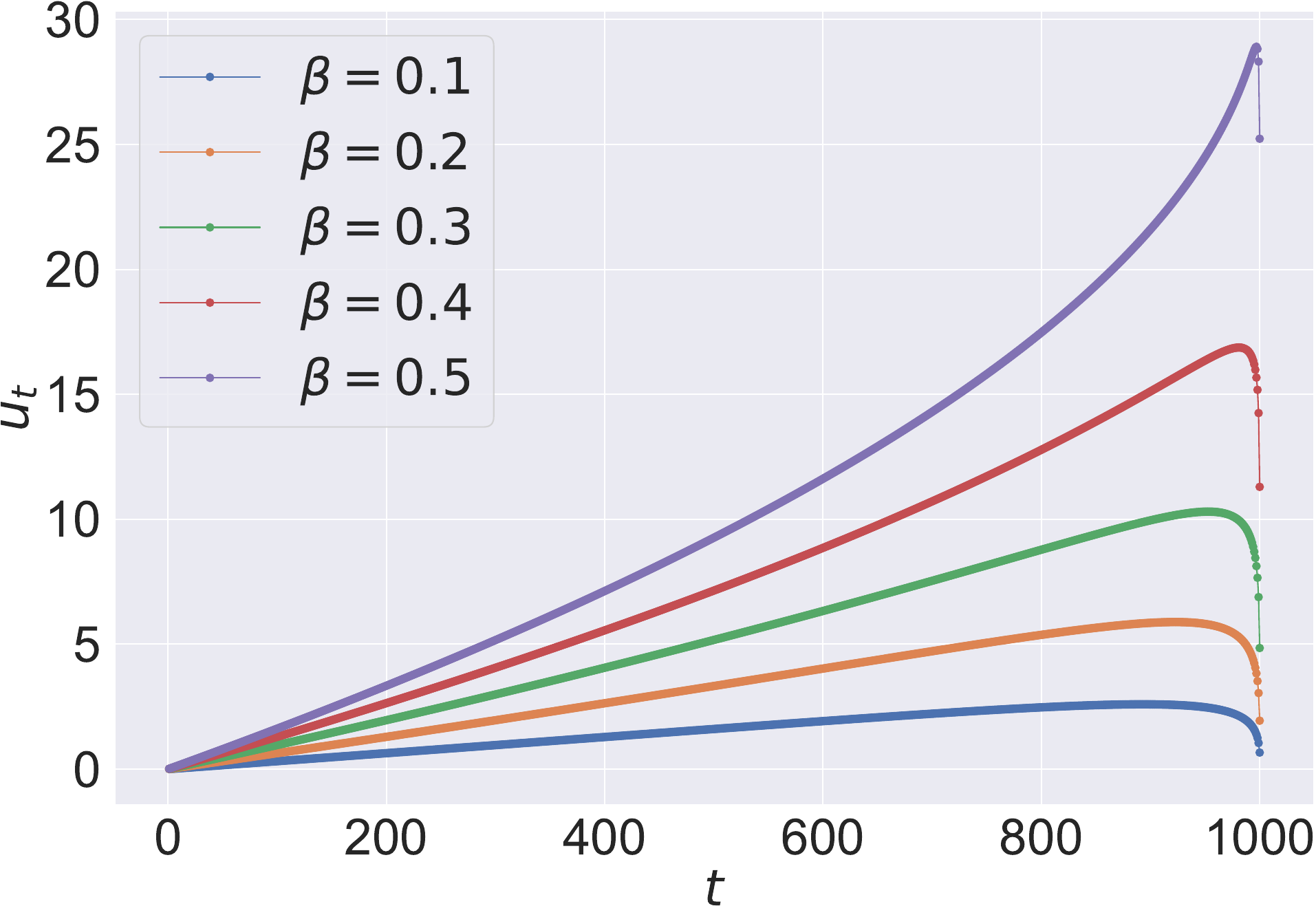}}\par\bigskip%
  \subcaptionbox{$\alpha = 10$, $T = 10$}
  [0.33\linewidth]{\includegraphics[width=\linewidth]{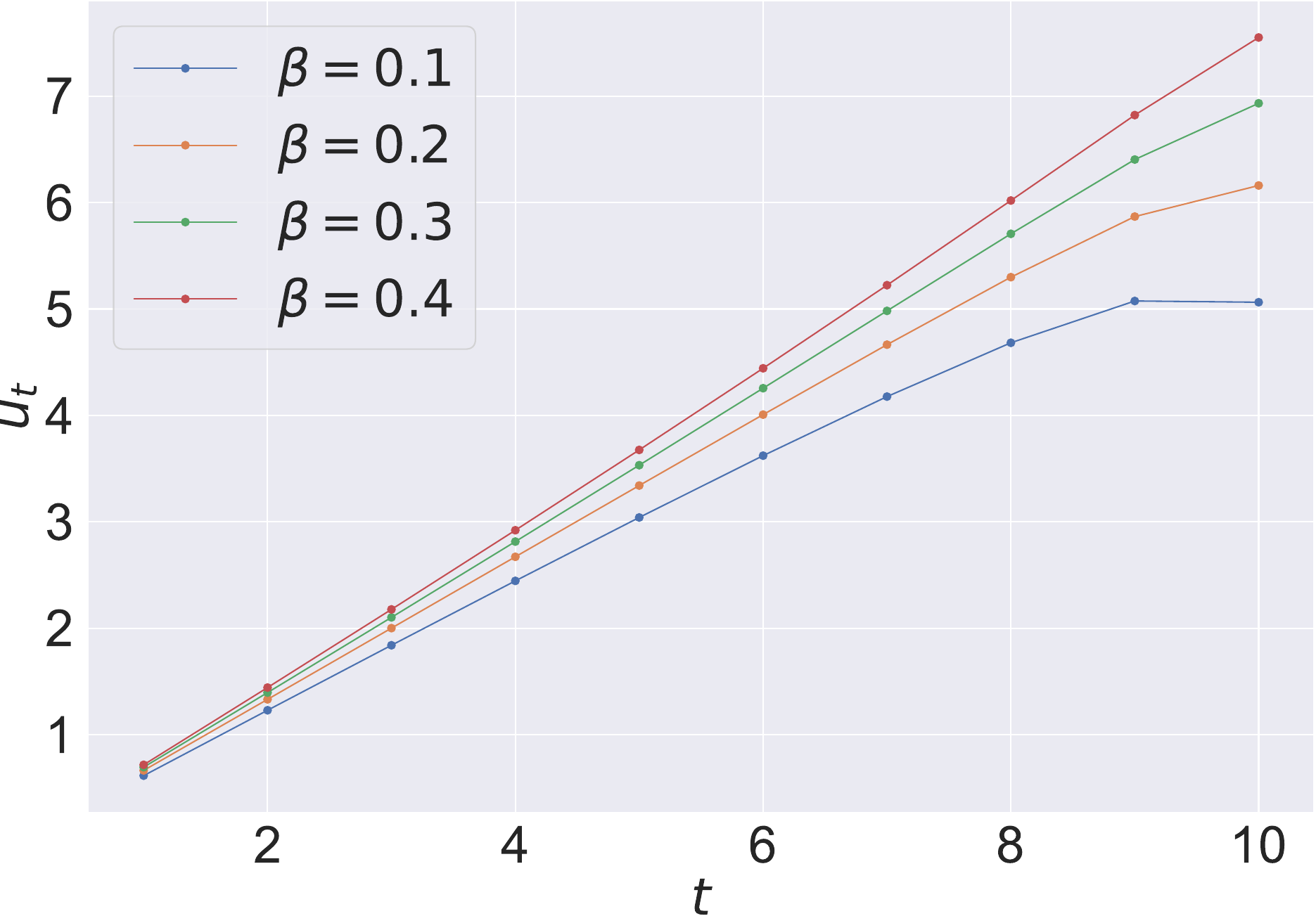}}\hfill%
  \subcaptionbox{$\alpha = 10$, $T = 100$}
  [0.33\linewidth]{\includegraphics[width=\linewidth]{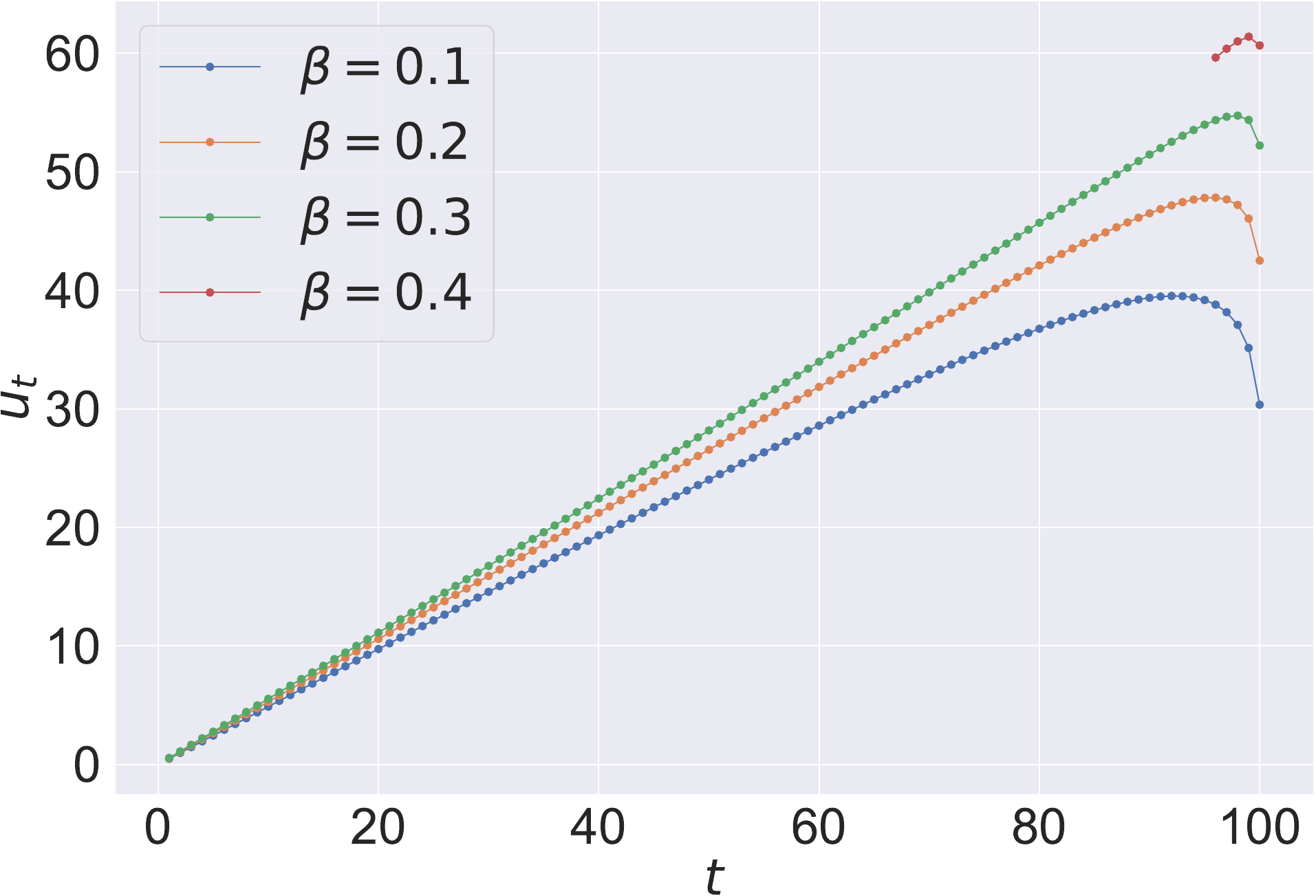}}\hfill%
  \subcaptionbox{$\alpha = 10$, $T = 1000$}
  [0.33\linewidth]{\includegraphics[width=\linewidth]{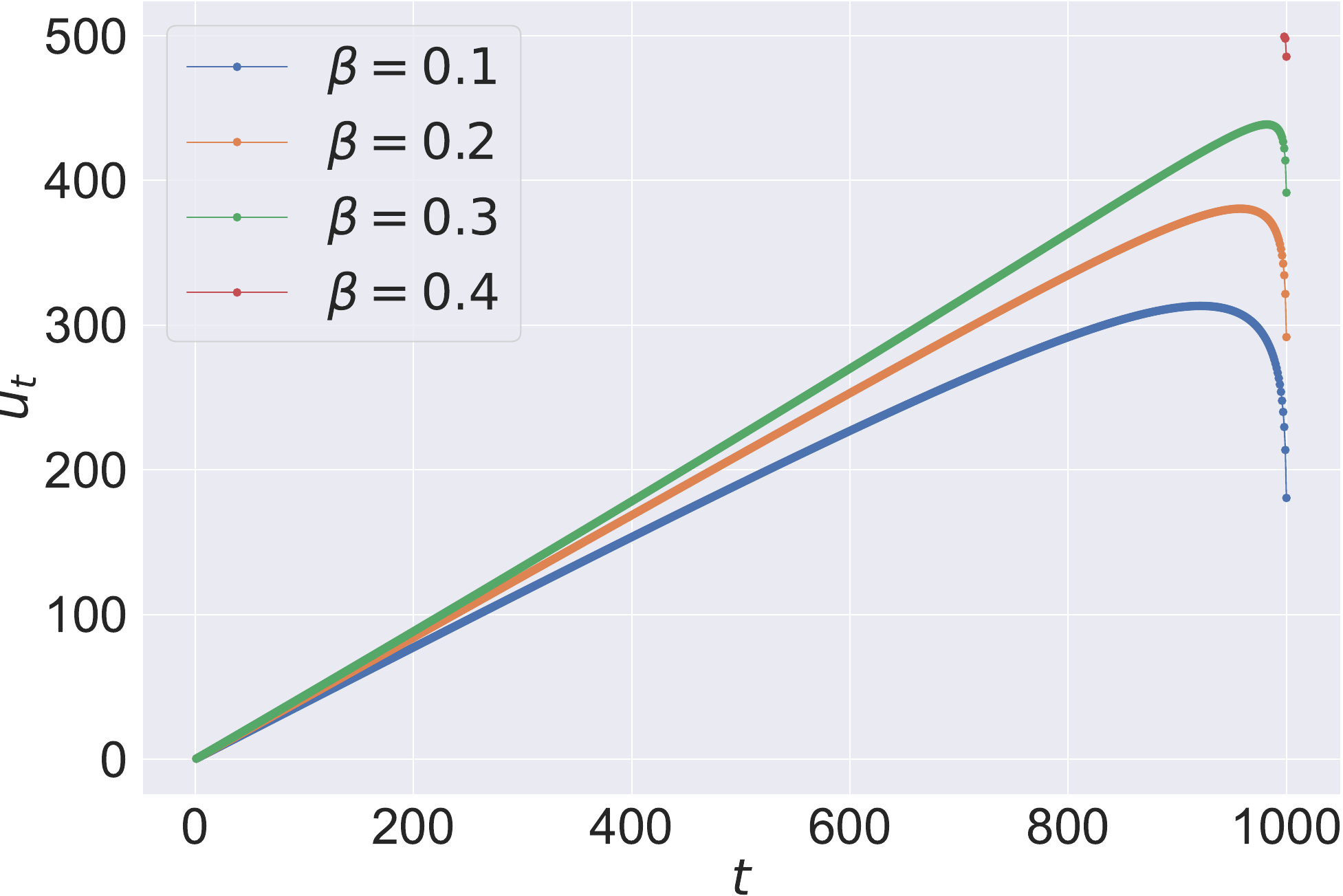}}\par%
  \caption{The plots of $u_t\ (t=\tilde{t}, \ldots, T)$ defined by \cref{eq:def_ut} for various $\alpha$, $\beta$, and $T$ when $R = 1$. We do not include cases where $\alpha = 10$ and $\beta = 0.5$, because $\beta_0 < 0.5$.}
  \label{fig:ut}
\end{figure}

\cref{fig:ut} shows $u_t$ for $t=\tilde{t}, \ldots, T$ defined by \cref{eq:def_ut} when $R=1$.
Let $t^* = \argmax_{t \in \set{\tilde{t}, \ldots, T}}u_t$. 
Since $u_t$ represents the maximum final progress when giving up at time $t$, it is optimal not to use an exploitative reward when $t^* = T$, and optimal to use an exploitative reward when $t^* \neq T$.
\cref{fig:ut} gives several insights about exploitative rewards. 
\begin{itemize}
    \item Using an exploitative reward tends to be optimal when $\alpha$ is large, $\beta$ is small, and $T$ is large. 
    \item When it is optimal to use an exploitative reward, $u_{t^*}/u_T$ becomes large when $\alpha$ is large, $\beta$ is small, and $T$ is large. Because $u_{t^*}$ and $u_{T}$ are the maximum values of final progress with and without exploitative rewards respectively, the larger $u_{t^*}/u_T$ is, the greater the effect of the exploitative rewards.
\end{itemize}
Exploitative rewards boost the final progress of agents with strong present bias; they get lured easily by exploitative rewards.
This tendency is significant for longer-term projects, suggesting that people with strong present bias are not good at envisioning the distant future.

The ineffectiveness of exploitative rewards with a small $\alpha$ can be understood as follows. 
When $\alpha$ is small, it is easy to make substantial progress in a single time step, so the agent makes little progress until near the end and only tries hard only at the end. 
As a result, the final progress will be small if the agent gives up in the middle, leading to the ineffectiveness of exploitative rewards.

\section{Optimal Reward Scheduling}
\label{sec:optimal scheduling}
Next, we consider a reward scheduling problem that aims to maximize the agent’s progress when the reward can be presented multiple times.
Our analysis will show that presenting rewards multiple times can increase the sum of progress compared to presenting all rewards at once.
We will also see that optimal reward scheduling varies greatly depending on the present-bias parameter $\beta$.

\begin{figure}[t]
  \centering
  \subcaptionbox{$k = 1$}
  [0.4\linewidth]{\includegraphics[width=\linewidth]{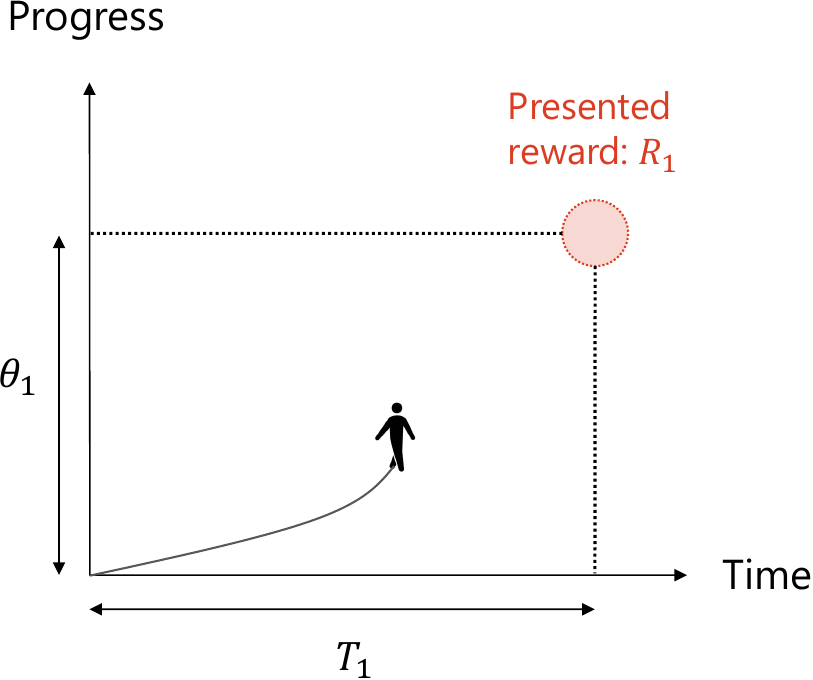}}\qquad\qquad%
  \subcaptionbox{$k = 3$}
  [0.4\linewidth]{\includegraphics[width=\linewidth]{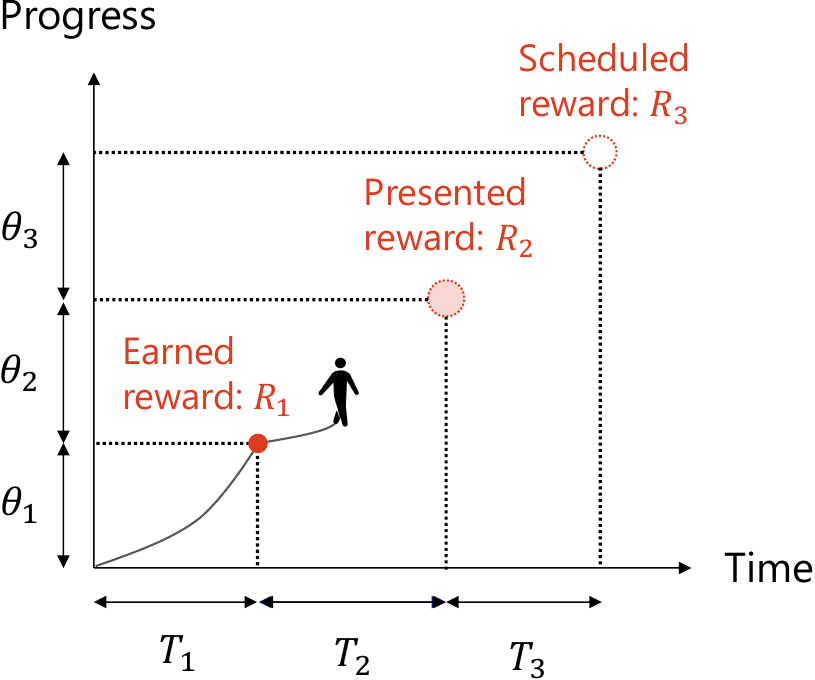}}%
  \caption{Examples of reward scheduling.}
  \label{fig:reward schedule}
\end{figure}

For the problem setup, we focus on the setting where exploitative rewards are not allowed to maintain the agent's motivation.
Given the total period $T \in \Z_{>0}$ and the total reward $R \in \R_{\geq 0}$, we divide them into $k$ periods $T_1,\dots,T_k \in \Z_{>0}$ and $k$ rewards $R_1,\dots,R_k \in \R_{\geq 0}$, where $\sum_{i=1}^k T_i = T$, $\sum_{i=1}^k R_i = R$, and $k \in \set{1,\dots,T}$ is arbitrary.
At the beginning of the $i$-th period, the agent is offered the reward $R_i$ and works toward the reward.
If the agent increases the progress by $\theta_i \in \R_{\geq 0}$ during the period, it receives the reward $R_i$.
Our goal is to maximize the sum of progress over the period under the constraint that the agent earns all rewards; this constraint derives from the setting where exploitative rewards are not allowed.
For this purpose, we seek optimal reward scheduling: $k$, $(T_i)_{i=1}^k$, $(R_i)_{i=1}^k$, and $(\theta_i)_{i=1}^k$.
\cref{fig:reward schedule} illustrates examples of reward scheduling in this setting.

\subsection{Non-TAI $\beta$}
From \cref{thm:goal_setting1}, the optimal $\theta_i$ is
\begin{align}
  \theta_i
  =
  R_i^{\frac{1}{\alpha}}
  \prn*{T_i - 1 + \beta^{\frac{1}{\alpha - 1}}}^{\frac{\alpha - 1}{\alpha}},
  \label{eq:optimal_theta_given_RT1}
\end{align}
given $R_i$ and $T_i$.
Thus, the reward scheduling problem is reduced to
\begin{align}
  \max_{k,\,(T_i)_{i=1}^k,\, (R_i)_{i=1}^k}\ 
  \sum_{i=1}^k 
  R_i^{\frac{1}{\alpha}}
  \prn*{T_i - 1 + \beta^{\frac{1}{\alpha - 1}}}^{\frac{\alpha - 1}{\alpha}},
  \quad
  \st\quad
  \sum_{i=1}^k T_i = T,\quad
  \sum_{i=1}^k R_i = R.
  \label{eq:problem_reward_scheduling1}
\end{align}
Note that $T_i \in \Z_{>0}$ but $R_i \in \R_{\geq 0}$.

\begin{theorem}
  The optimal reward scheduling is
  \begin{align}
    k = 1,\quad
    T_1 = T,\quad
    R_1 = R,\quad
    \theta_1
    =
    R^{\frac{1}{\alpha}}
    \prn*{T - 1 + \beta^{\frac{1}{\alpha - 1}}}^{\frac{\alpha - 1}{\alpha}}.
  \end{align}
  Then the agent's progress is $R^{\frac{1}{\alpha}}
  \prn[\big]{T - 1 + \beta^{\frac{1}{\alpha - 1}}}^{\frac{\alpha - 1}{\alpha}}$.
\end{theorem}
\begin{proof}
  H\"older's inequality gives an upper bound on the objective function of problem~\cref{eq:problem_reward_scheduling1}:
  \begin{align}
    \sum_{i=1}^k
    R_i^{\frac{1}{\alpha}}
    \prn*{T_i - 1 + \beta^{\frac{1}{\alpha - 1}}}^{\frac{\alpha - 1}{\alpha}}
    &\leq
    \prn*{\sum_{i=1}^k R_i}^{\frac{1}{\alpha}}
    \prn*{\sum_{i=1}^k \prn*{T_i - 1 + \beta^{\frac{1}{\alpha - 1}}}}^{\frac{\alpha - 1}{\alpha}}\\
    &=
    R^{\frac{1}{\alpha}}
    \prn*{T - k \prn*{1 - \beta^{\frac{1}{\alpha - 1}}}}^{\frac{\alpha - 1}{\alpha}}
    \leq
    R^{\frac{1}{\alpha}}
    \prn*{T - 1 + \beta^{\frac{1}{\alpha - 1}}}^{\frac{\alpha - 1}{\alpha}},
  \end{align}
  and the upper bound is achieved when $k=1$, $T_1 = T$, and $R_1 = R$.
  Hence, the optimal solution is $k=1$, $T_1 = T$, and $R_1 = R$.
  The optimal $\theta$ is obtained from \cref{eq:optimal_theta_given_RT1}.
\end{proof}

\subsection{TAI $\beta$}
From \cref{thm:goal_setting2_nonexploitative}, the optimal $\theta_i$ is
\begin{align}
  \theta_i
  =
  (\beta R_i)^{\frac{1}{\alpha}}
  \frac{\Gamma(T_i + \beta^{\frac{1}{\alpha - 1}})}{\Gamma(T_i) \Gamma(1 + \beta^{\frac{1}{\alpha - 1}})},
  \label{eq:optimal_theta_given_RT2}
\end{align}
given $R_i$ and $T_i$.
Thus, the reward scheduling problem is reduced to
\begin{align}
  \max_{k,\,(T_i)_{i=1}^k,\, (R_i)_{i=1}^k}\ 
  \sum_{i=1}^k
  (\beta R_i)^{\frac{1}{\alpha}}
  \frac{\Gamma(T_i + \beta^{\frac{1}{\alpha - 1}})}{\Gamma(T_i) \Gamma(1 + \beta^{\frac{1}{\alpha - 1}})},\quad
  \st\quad
  \sum_{i=1}^k T_i = T,\quad
  \sum_{i=1}^k R_i = R.
  \label{eq:problem_reward_scheduling2}
\end{align}

Let us first fix $k$ and $(T_i)_{i=1}^k$.
Then the optimal $(R_i)_{i=1}^k$ is obtained as follows:
H\"older's inequality gives
\begin{align}
  \sum_{i=1}^k
  R_i^{\frac{1}{\alpha}}
  \frac{\Gamma(T_i + \beta^{\frac{1}{\alpha - 1}})}{\Gamma(T_i)} 
  &\leq
  \prn*{\sum_{i=1}^k R_i}^{\frac{1}{\alpha}}
  \prn*{\sum_{i=1}^k \prn[\bigg]{\frac{\Gamma(T_i + \beta^{\frac{1}{\alpha - 1}})}{\Gamma(T_i)}}^{\frac{\alpha}{\alpha - 1}}}^{\frac{\alpha - 1}{\alpha}}
  =
  R^{\frac{1}{\alpha}}
  \prn*{\sum_{i=1}^k F(T_i)}^{\frac{\alpha - 1}{\alpha}},
\end{align}
where
\begin{align}
  F(x)
  \coloneqq
  \prn[\bigg]{ \frac{\Gamma(x + \beta^{\frac{1}{\alpha - 1}})}{\Gamma(x)} }^{\frac{\alpha}{\alpha - 1}},
  \label{eq:def_F}
\end{align}
and the equality holds when
\begin{align}
  R_i
  &=
  R
  \frac{F(T_i)}{\sum_{j=1}^k F(T_j)}  
  \label{eq:optimal_R_given_T}
\end{align}
for all $1 \leq i \leq k$.

Next, let us optimize $k$ and $(T_i)_{i=1}^k$.
Problem~\cref{eq:problem_reward_scheduling2} is now reduced to
\begin{align}
  \max_{k,\,(T_i)_{i=1}^k}\ 
  \sum_{i=1}^k F(T_i),\quad
  \st\quad
  \sum_{i=1}^k T_i = T.
  \label{eq:problem_reward_scheduling2_reduced}
\end{align}
Although this problem is difficult to solve explicitly, we can compute the optimal solution by the following algorithmic procedure.

Let $v_T$ be the optimal value of problem~\cref{eq:problem_reward_scheduling2_reduced} and let $v_0 \coloneqq 0$.
Considering the case of $T_1 = 1,2,\dots,T$ separately yields a recursive formula:
\begin{align}
  v_T
  =
  \max_{1 \leq t \leq T}
  \set*{ F(t) + v_{T-t} }.
  \label{eq:reward_scheduling_dp}
\end{align}
After computing $v_1,\dots,v_T$ according to this formula, find $t \in \set{1,\dots,T}$ such that $v_T = F(t) + v_{T-t}$.
Such $t$ implies that $T_1 = t$ in an optimal solution to problem~\cref{eq:problem_reward_scheduling2_reduced}, and thus the problem size is reduced from $T$ to $T-t$.
Repeating this reduction gives us the optimal solution to problem~\cref{eq:problem_reward_scheduling2_reduced}.
The above procedure can be performed in $O(T^2)$ time.

We can also observe the effect of splitting the reward by examining the two extreme cases: offering all rewards at once, i.e.,
\begin{align}
  k = 1,\quad
  T_1 = T,\quad
  R_1 = R,\quad
  \theta_1
  =
  (\beta R)^{\frac{1}{\alpha}}
  \frac{\Gamma(T + \beta^{\frac{1}{\alpha - 1}})}{\Gamma(T) \Gamma(1 + \beta^{\frac{1}{\alpha - 1}})},
\end{align}
and offering the reward every time, i.e.,
\begin{align}
  k = T,\quad
  T_1 = \dots = T_k = 1,\quad
  R_1 = \dots = R_k = \frac{R}{T},\quad
  \theta_1 = \dots = \theta_k = \prn*{\frac{\beta R}{T}}^{\frac{1}{\alpha}}.
\end{align}
The ratio of progress achieved in each case is
\begin{align}
  T \prn*{\frac{\beta R}{T}}^{\frac{1}{\alpha}} \bigg/
  \prn[\bigg]{
    (\beta R)^{\frac{1}{\alpha}}
    \frac{\Gamma(T + \beta^{\frac{1}{\alpha - 1}})}{\Gamma(T) \Gamma(1 + \beta^{\frac{1}{\alpha - 1}})}
  }
  &=
  T^{1 - \frac{1}{\alpha}}
  \frac{\Gamma(T) \Gamma(1 + \beta^{\frac{1}{\alpha - 1}})}{\Gamma(T + \beta^{\frac{1}{\alpha - 1}})}\\
  &=
  \Gamma(1 + \beta^{\frac{1}{\alpha - 1}})
  T^{1 - \frac{1}{\alpha} - \beta^{\frac{1}{\alpha - 1}}}
  + O \prn[\big]{ T^{- \frac{1}{\alpha} - \beta^{\frac{1}{\alpha - 1}}} }
\end{align}
as $T \to \infty$, where we have used the asymptotic formula \citep[Eq.~(1)]{tricomi1951asymptotic}.
If the present-bias parameter $\beta$ is as small as $\beta \leq (1 - \frac{1}{\alpha})^{\alpha - 1}$, this ratio diverges to infinity as $T \to \infty$.
The smaller $\beta$, the faster the divergence.
The results suggest that frequent rewards are effective for agents with strong present bias.

\subsection{TAI $\beta$ and $\alpha = 2$}
Assuming $\alpha = 2$ enables us to analyze problem~\cref{eq:problem_reward_scheduling2_reduced} in more detail.
For convenience, let $F(0) \coloneqq \lim_{x \searrow 0} F(x) = 0$.

\begin{theorem}
  Suppose that $\alpha = 2$.
  \leavevmode
  \begin{enuminthm}
    \item
    \label{thm:reward_scheduling1}
    If $\frac{1}{2} \leq \beta < \beta_0$, the optimal reward scheduling is
    \begin{align}
      k = 1,\quad
      T_1 = T,\quad
      R_1 = R,\quad
      \theta_1 = \sqrt{\beta R} \frac{\Gamma(T + \beta)}{\Gamma(T) \Gamma(1 + \beta)}.
    \end{align}
    Then the agent's progress is $\sqrt{\beta R} \frac{\Gamma(T + \beta)}{\Gamma(T) \Gamma(1 + \beta)}$.
    \item
    \label{thm:reward_scheduling2}
    If $\beta \leq \sqrt{2} - 1$, the optimal reward scheduling is
    \begin{align}
      k = T,\quad
      T_1 = \dots = T_k = 1,\quad
      R_1 = \dots = R_k = \frac{R}{T},\quad
      \theta_1 = \dots = \theta_k = \sqrt{\frac{\beta R}{T}}.
    \end{align}
    Then the agent's progress is $\sqrt{\beta R T}$.
  \end{enuminthm}
\end{theorem}
\begin{proof}[Proof of \cref{thm:reward_scheduling1}]
  Since
  \begin{align}
    \frac{F(x+2) + F(x)}{F(x+1)} - 2
    &=
    \prn*{
      \frac{\Gamma(x + \beta + 2) \Gamma(x + 1)}{\Gamma(x + \beta + 1) \Gamma(x + 2)}
    }^2
    + 
    \prn*{
      \frac{\Gamma(x + \beta) \Gamma(x + 1)}{\Gamma(x + \beta + 1) \Gamma(x)}
    }^2
    - 2\\
    &=
    \prn*{ \frac{x + \beta + 1}{x + 1} }^2
    + \prn*{ \frac{x}{x + \beta} }^2
    - 2\\
    &=
    \beta \frac{(1 - \beta)^2 + (2 \beta - 1) (2x + \beta + 1)^2}{2 (x + 1)^2 (x + \beta)^2}
    \geq 0,
  \end{align}
  we have $F(x+2) - F(x+1) \geq F(x+1) - F(x)$ for all $x \in \Z_{\geq 0}$.
  Applying this inequality repeatedly yields 
  \begin{align}
    F(x + y) - F(x)
    =
    \sum_{i=x}^{x+y-1} \prn*{ F(i+1) - F(i) }
    \geq
    \sum_{i=0}^{y-1} \prn*{ F(i+1) - F(i) }
    =
    F(y) - F(0)
    =
    F(y),
  \end{align}
  and thus $F(x) + F(y) \leq F(x + y)$ for any $x, y \in \Z_{\geq 0}$.
  This implies an upper bound on the objective function of problem~\cref{eq:problem_reward_scheduling2_reduced}: $\sum_{i=1}^k F(T_i) \leq F(T)$, and the upper bound is achieved when $k = 1$ and $T_1 = T$.
  Hence, the optimal solution to problem~\cref{eq:problem_reward_scheduling2_reduced} is $k = 1$ and $T_1 = T$.
  The optimal $\theta_i$ and $R_i$ are obtained from \cref{eq:optimal_theta_given_RT2,eq:optimal_R_given_T}.
\end{proof}
\begin{proof}[Proof of \cref{thm:reward_scheduling2}]
  Let $G(x) \coloneqq F(x) / x$.
  Since
  \begin{align}
    \frac{G(x+1)}{G(x)} - 1
    &=
    \prn*{
      \frac{\Gamma(x + \beta + 1) \Gamma(x)}{\Gamma(x + \beta) \Gamma(x + 1)}
    }^2
    \frac{x}{x+1}
    - 1
    =
    \prn*{ \frac{x + \beta}{x} }^2
    \frac{x}{x+1}
    - 1
    =
    \frac{\beta^2 - (1 - 2 \beta) x}{x (x + 1)}
    \label{eq:Gx_ratio}
  \end{align}
  and $1 - 2 \beta > 0$, we have $G(x+1) \leq G(x)$ for $x \geq \frac{\beta^2}{1 - 2 \beta}$.
  Here, we also have $\frac{\beta^2}{1 - 2 \beta} \leq 1$ since $\beta \leq \sqrt{2} - 1$.
  Therefore, $G(1) \geq G(2) \geq \cdots$ holds.
  Using this inequality yields an upper bound on the objective function of problem~\cref{eq:problem_reward_scheduling2_reduced}:
  \begin{align}
    \sum_{i=1}^k F(T_i)
    =
    \sum_{i=1}^k T_i G(T_i)
    \leq
    \sum_{i=1}^k T_i G(1)
    =
    T G(1),
  \end{align}
  and the upper bound is achieved when $k = T$ and $T_1 = \cdots = T_k = 1$.
  Hence, the optimal solution to problem~\cref{eq:problem_reward_scheduling2_reduced} is $k = T$ and $T_1 = \dots = T_k = 1$.
  The optimal $\theta_i$ and $R_i$ recover from \cref{eq:optimal_theta_given_RT2,eq:optimal_R_given_T}.
\end{proof}

\begin{figure}
  \centering
  \subcaptionbox{$T = 10$}
  [0.6\linewidth]{\includegraphics[width=\linewidth]{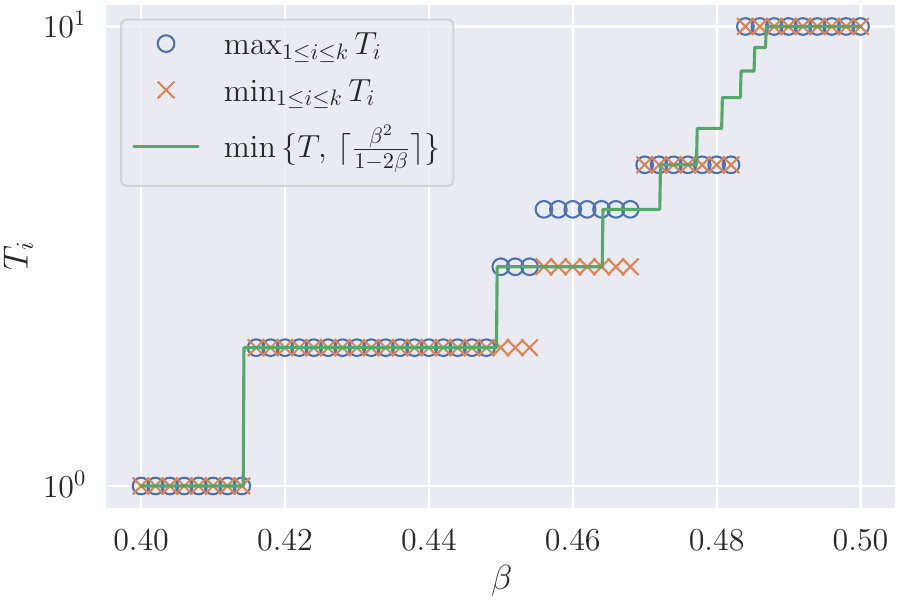}}\par\bigskip%
  \subcaptionbox{$T = 1000$}
  [0.6\linewidth]{\includegraphics[width=\linewidth]{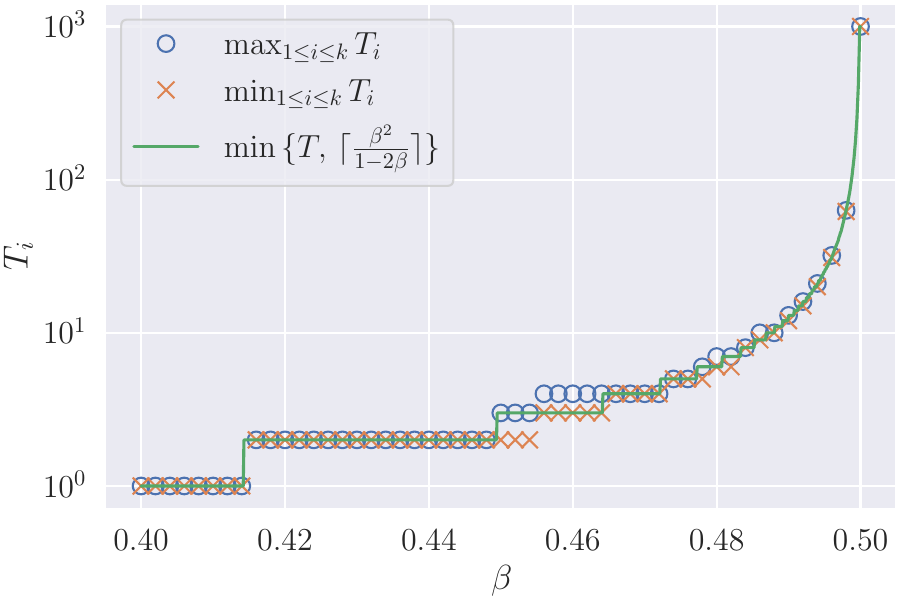}}%
  \caption{
    Optimal reward interval with $\alpha = 2$.
    The markers indicate the maximum and minimum lengths of the periods in the optimal solution $(T_i)_{i=1}^k$ computed with \cref{eq:reward_scheduling_dp}.
    The green line shows the nearly optimal reward interval \cref{eq:nearly_optimal_reward_interval} derived from the theoretical analysis.
  }
  \label{fig:optimal_reward_interval}
\end{figure}

The case of $\sqrt{2} - 1 < \beta < \frac{1}{2}$ is the most nontrivial.
Let us consider $G(x) = F(x) / x$ introduced in the proof of \cref{thm:reward_scheduling2}.
The value $G(x)$ represents the increase in the objective function value per unit of time when the reward is placed at time $x$.
Intuitively, placing a reward every $x^*$ units of time is expected to be nearly optimal, where $x^*$ is $x \in \set{1, \dots, T}$ that maximizes $G(x)$, and we have
\begin{align}
  x^*
  =
  \min \set*{T,\ \ceil*{\frac{\beta^2}{1 - 2 \beta}}}
  \label{eq:nearly_optimal_reward_interval}
\end{align}
from \cref{eq:Gx_ratio}.
\Cref{fig:optimal_reward_interval} compares the nearly optimal reward interval \cref{eq:nearly_optimal_reward_interval} with the optimal reward interval computed with \cref{eq:reward_scheduling_dp}.
This suggests that the analytical approximate solution~\cref{eq:nearly_optimal_reward_interval} agrees with the exact solution when $\beta$ is relatively small or $T$ is large.

\subsection{Discussion}
The results indicate that when the present-bias parameter $\beta$ is large, the agent's progress can be maximized by giving the reward in a lump sum, but when $\beta$ is small, giving the reward multiple times enhances the progress.
This is because agents with strong present bias are more likely to abandon a task due to procrastination.
Appropriate intermediate rewards can prevent them from abandoning the task and increase their progress.
Although there have been studies on the algorithmic aspects of intermediate rewards, no studies have quantitatively discussed the relationship between the strength of present bias and the effectiveness of intermediate rewards.
This finding is also significant when using incentives in the real world.
It suggests that it is desirable to tailor the reward plan to each person's present bias, rather than providing rewards uniformly to everyone.

\section{Conclusion and Future Directions}
In this paper, we proposed a model for understanding the goal-achieving behavior of agents under the influence of present bias.
Although our model is restricted to tasks that accumulate progress, the agent's behavior can be solved analytically, which is the key feature. 
Based on the analytical solution, we analyzed three problems: task abandonment, goal optimization problem, and reward scheduling problem.
We obtained several insights into the relationship between the strength of present bias and these problems and their application in the real world.

The following three extensions to our study should be considered in the future.
\begin{itemize}
  \item
  \emph{Additional biases.}
  To further understand human behavior, it is necessary to consider the influence of biases other than present bias, such as sunk-cost bias \cite{arkes1985psychology} and loss aversion \cite{tversky1992advances}.
  In addition, \emph{naivete and sophistication} have been studied in relation to procrastination \cite{o1999doing, o2001choice}. We analyzed the behavior of naive people who are unaware of the influence of present bias, but it is known that there are sophisticated people conscious of it. 
  It is challenging but vital to include the effects of these various biases in the model while still maintaining its ease of use.
  \item
  \emph{Types of tasks.}
  This study assumed progress to increase monotonically, but this may not be true for some tasks.
  For example, if a dieter defines progress as the amount of weight loss, progress can decrease.
  It is worthwhile to extend our model to handle a broader range of real-world tasks while maintaining tractability.
  \item
  \emph{Multiple simultaneous rewards.}
  In \cref{sec:optimal scheduling}, we assumed that each reward is offered in order.
  However, it may be possible to motivate agents more efficiently by presenting multiple rewards simultaneously.
  Multiple rewards complicate the agent's behavior, but if it can be handled analytically, the range of intervention will be significantly expanded.
\end{itemize}

\bibliographystyle{abbrvnat}
\bibliography{main}

\end{document}